\documentstyle[psfig,11pt,aaspp4]{article}

\def\ale{\mathrel{\hbox{\rlap{\hbox{\lower4pt\hbox{$\sim$}}}\hbox{$<$}}}}
\def\age{\mathrel{\hbox{\rlap{\hbox{\lower4pt\hbox{$\sim$}}}\hbox{$>$}}}}

\def\arcsec{\hbox{$^{\prime\prime}$}}

\def\gsim{\mathrel{\hbox{\rlap{\lower.55ex \hbox {$\sim$}}
                   \kern-.3em \raise.4ex \hbox{$>$}}}}
\def\lsim{\mathrel{\hbox{\rlap{\lower.55ex \hbox {$\sim$}}
                   \kern-.3em \raise.4ex \hbox{$<$}}}}
\def\grb{GRB\thinspace{970508}}

\def\aa#1#2#3{\bibitem[]{}#1, A\&A, #2, #3.}

\def\aj#1#2#3{\bibitem[]{}#1, AJ, #2, #3.}
\def\apj#1#2#3{\bibitem[]{}#1, {\it Ap. J.}, {\bf#2}, #3.}
\def\apjlett#1#2#3{\bibitem[]{}#1, {\it Ap. J. (Letters)}, {\bf #2},
#3.}

\def\iauc#1#2{\bibitem[]{}#1, IAU Circ.~No.~#2}

\def\mnras#1#2#3{\bibitem[]{}#1, {\it M.N.R.A.S.}, {\bf#2}, #3.}
\def\nature#1#2#3{\bibitem[]{}#1, {\it Nature}, {\bf #2}, #3.}

\def\science#1#2#3{\bibitem[]{}#1, Science, #2, #3.}

\lefthead{Frail et al.}
\righthead{The Radio Afterglow of GRB 970508}
\begin{document}

\title{A 450-day Light Curve of the Radio Afterglow of GRB 970508: 
Fireball Calorimetry}

\author{D.~A.~Frail\altaffilmark{1}, E. Waxman\altaffilmark{2,3},
\&  S.~R.~Kulkarni\altaffilmark{4}}

\altaffiltext{1}{National Radio Astronomy Observatory, P.~O.~Box O,
  Socorro, NM 87801, USA}

\altaffiltext{2}{Institute for Advanced Study, Princeton, NJ 08540, USA}

\altaffiltext{3}{Department of Condensed-Matter Physics, Weizmann
  Institute, Rehovot 76100, Israel}

\altaffiltext{4}{California Institute of Technology, Owens Valley Radio
Observatory,
105-24, Pasadena, CA 91125, USA}

\begin{abstract}

  We report on the results of an extensive monitoring campaign of the
  radio afterglow of GRB 970508, lasting 450 days after the burst. The
  spectral and temporal radio behavior indicate that the fireball has
  undergone a transition to sub-relativistic expansion at $t\sim100$
  days. This allows us to perform ``calorimetry'' of the explosion.
  The derived total energy, $E_0\sim 5\times 10^{50}$ erg is well
  below the $\sim 5\times10^{51}$ erg inferred under the assumption
  of spherical symmetry from $\gamma$-ray and early afterglow
  observations. A natural consequence of this result, which can also
  account for deviations at $t<100$ days from the spherical
  relativistic fireball model predictions, is that the fireball was
  initially a wide-angle jet of opening angle $\sim 30^\circ$.

Our analysis also allows to determine the energy fractions carried by
electrons and magnetic field, and the density of ambient medium
surrounding the fireball.  We find that during the sub-relativistic
expansion electrons and magnetic field are close to equipartition, and
that the density of the ambient medium is $\sim~1\, {\rm cm}^{-3}$. The
inferred density rules out the possibility that the fireball expands
into a strongly non-uniform medium, as would be expected, e.g., in the
case of a massive star progenitor.

\end{abstract}
\clearpage
\section{Introduction}\label{sec:introduction}

The gamma-ray burst (GRB) of May 8, 1997 was a watershed event in the
study of these intriguing objects. There were two major advances
resulting from afterglow observations of \grb: (1) The first
unambiguous evidence that some GRBs are a cosmological population was
obtained with the detection of red-shifted metal lines in absorption
against the optical afterglow (Metzger et al. 1997). The resulting
lower limit on the distance, combined with the measured gamma-ray
fluence, established an approximate energy scale for these events. (2)
The discovery of the first radio afterglow and the demonstration
of relativistic expansion of the fireball (Frail et al. 1997).
The latter was made possible by the observations of distinctive
variations in the radio flux
and attributed to interstellar scintillation (Goodman
1997).
It is worth noting that GRBs now join
quasars and Galactic micro-quasars as sources for which superluminal
motions have been inferred.

The afterglow from GRB 970508 was particularly bright and long-lived.
Indeed, to this date, this GRB remains  unusual in both these
respects. Consequently,  astronomers carried out
intensive observations of the
afterglow across the electromagnetic spectrum.  The principal
motivation of the observations were to infer the fundamental
parameters of the explosion: the total energy of the explosion
($E_0$), the distribution of the circumburst medium (density, $n_i$
and possible radial dependence as would be the case if the explosion
took place in a circumburst medium shaped by mass loss from
the progenitor) and the geometry of
the afterglow (sphere versus collimated flow or ``jets'').

Of all the parameters listed above, $E_0$ is perhaps the most eagerly
sought parameter. After all, in a manner similar to supernovae, it is
$E_0$ which sets the scale for the entire GRB phenomenon and thus it
is a truly fundamental parameter of the explosion.  With few
exceptions (Katz \& Piran 1997, Brainerd 1998) most of the early
estimates for $E_0$ (Waxman 1997, Wijers \& Galama 1999, Granot et al.
1999) lie within a factor of three of $10^{52}$ erg. Although their
precise methodologies differ, all of these determinations use
information gleaned from the early afterglow observations.
Unfortunately,
the Lorentz factor is large during the early phase and due to
relativistic beaming the observer sees only solid angle $\Gamma^{-2}$
of the emitting surface. Thus
the inferred energy is necessarily the
``isotropic equivalent'' value. However, if
fireballs are jet sources (as appears to be the case), such
``isotropic equivalent'' estimates are upper limits and
in some cases can be gross upper limits (e.g. GRB 990510,
Harrison et al. 1999)

Radio observations are essentially immune from the geometry of the
fireball and thus offer us the best method to infer $E_0$ since much of
the radio afterglow is emitted at later epoch when $\Gamma$ is
falling.  Indeed, over the duration of the radio emission, the emitting
material typically becomes sub-relativistic.  Once the flow becomes
sub-relativistic it will in due course also become spherical. Thus
observations of radio afterglows allow us the opportunity of inferring
$E_0$ without the usual concerns of the geometry of the fireball.

Radio observations offer two additional advantages: (1) At X-ray and
optical wavelengths, the afterglow emission rapidly enters the
monotonic decaying regime within minutes to hours of the burst.
Unfortunately, at the current time, logistical difficulties prevent us
from responding so rapidly to GRB events. X-ray and optical afterglow
observations are typically initiated hours after the burst.  However,
at radio wavelengths, the entire afterglow phenomenon is stretched in
time. Thus monitoring the radio afterglow observations allows us to
see all the important transitions (save the cooling transition) which
in turn provides key diagnostics to infer the fundamental parameters
of the explosion. 
(2) Radio observations offer us,
via interstellar scintillations, the only way to measure the tiny
angular size (few microarcseconds) of GRB afterglows. The measured
size in turn allows us to verify the dynamics of the explosion.

In this paper, we present the complete radio light curves from the
radio afterglow of \grb\ beginning 3.5 hrs after the burst and ending
450 d later when the source was no longer detectable. This is the
third and final paper of an ambitious monitoring program that we
initiated following the discovery of radio afterglow from \grb.  In
the first paper (Frail et al.~1997) we reported the discovery of the
radio afterglow and the first 90 d of observations, and we interpreted
the strong variations as being due to diffractive scintillation. In
the second paper (Waxman, Kulkarni \&\ Frail 1998; hereafter, WKF98) 
we compared these
observations to theoretical models and showed that the size estimated
from diffractive scintillation was in excellent accord with that
expected from hydrodynamical models. In the same paper we 
noted that the observed flux at late times ($t>25$ d)
was well below that predicted by a spherical model and suggested
that this deviation was most likely
explained by a jet-like geometry for the fireball rather than a
sphere. We suggested that $E_0$ would be significantly
lower than the $10^{52}$ erg isotropic equivalent estimate from
the early afterglow observations and the isotropic equivalent
$\gamma$-ray energy release and further suggested that the fireball
would undergo a transition from relativistic to sub-relativistic expansion
at $t\sim100$~d.

The organization of the paper is as follows.  The observations are
summarized in \S\ref{sec:observations}. In \S\ref{sec:results} we
present the basic results in the form of
light curves in three bands, 8.46 GHz,
4.86 GHz and 1.43 GHz.
In \S\ref{sec:riss} we note a transition
of the variations in the light curve from the diffractive to the
refractive regimes. In \S\ref{sec:Spherical} we compare the 
data to the expectations from the
simple adiabatic fireball model.
In \S\ref{sec:Deviations} we
point out important deviations from model predictions and argue
that the afterglow was not spherical but was a jet with an
opening angle of about 30 degrees. Finally, in
\S\ref{sec:sub-Relativistic} we propose a detailed sub-relativistic
model which provides a satisfactory explanation to the very late
time observations. The transition to non-relativistic regime
allows us to carry out proper calorimetry of the explosion 
(\S\ref{sec:Calorimetry}).

\section{Observations}\label{sec:observations}

Except where noted, all observations reported here were obtained at
the Very Large Array (VLA) of the National Radio Astronomy Observatory
(NRAO).\footnotemark\footnotetext{The NRAO is a facility of the
  National Science Foundation operated under cooperative agreement by
  Associated Universities, Inc.} The very first observation was made
in the 1.4-GHz band, only 3.5 hrs after the initial gamma-ray
detection of \grb\ (Costa et al. 1997). Table 1 lists each observing
session giving the starting date, the time elapsed since the burst,
the 1.43 GHz flux density of \grb\ (F$_{1.4}\pm\sigma_{1.4}$) and the
VLA configuration. The first detection of the source was made on 1997
May 13.96 UT at 8.46 GHz.  Except for the first week, it was customary
to observe at 8.46 GHz and 4.86 GHz simultaneously, so we include them
together in Table 2.  All flux density measurements prior to 1997
August were previously compiled in Frail et al.~(1997) but are listed
here for completeness.

Calibration of the array phase was accomplished using extragalactic
radio sources with well-known positions. J0410+769 was used primarily
at 1.43 GHz while J0726+791 was the calibrator of choice for 4.86 GHz
and 8.46 GHz.  Since these compact sources usually vary on time scales
of a month or more, the absolute flux scale (accurate to better than
$\pm$2\%) at each epoch for all frequencies was fixed with
observations of 3C\thinspace{48}, 3C\thinspace{147}, and/or
3C\thinspace{286}.

The off-line analysis of the VLA data proceeded following standard
practice using the AIPS data reduction package. When the radio
afterglow from \grb\ was clearly detected, its flux density was
measured by a Gaussian fitting procedure which solves for the peak and
integrated intensity, the source position and its angular size.  When
the source is weak ($<5\sigma$) and no reliable fit was possible, the
peak flux density was measured from the image at the reported position
of \grb\ (Taylor et al.~1997).

The error estimates in Tables 1 and 2 are based on the thermal noise
fluctuations in the final image and are estimated by calculating the
rms fluctuations in a region free of any obvious radio sources. In the
D-array and to some extent in the CS array, where the resolution at
1.43 GHz is poor, the limiting sensitivity is not due to the receiver
noise but to the contribution of unresolved, faint sources within the
beam. This quantity varies from point to point in the sky but is of
order 80 $\mu$Jy for a 45\arcsec\ beam at 1.4 GHz (Condon et
al.~1998).  Improperly cleaned sidelobes from the myriad sources in
the D-array 1.43 GHz images constitute a further source of confusion.
Together these effects can amount to 100 $\mu$Jy of uncertainty added
in quadrature to the error estimates of the D-array measurements given
in Table 1.

While the source confusion discussed above is negligible at 4.86 and
8.46 GHz, there is a second source of error in the largest array
configurations (A and B) which can also be difficult to quantify.
Atmospheric phase instabilities occurring on a time scale shorter than
or comparable to the cycle time between calibrator and source will
``scatter'' flux from \grb\ and may introduce slight position offsets.
Large and erratic excursions of the calibrator phase over the course
of an observing run are a useful diagnostic to indicate when this is
occurring. Suspect data is revealed in off-line processing by sudden
jumps in the flux density of the phase calibrator from its long-term
average and a significant difference between the peak and integrated
flux densities in the Gaussian fits of the radio emission from \grb.
Following these criteria it was necessary to discard data from only
four epochs in 1997 (Sept. 19, Sept. 21, Dec. 23 and Dec. 26).

\section{Results: Light Curves and Spectra}\label{sec:results}
         
\begin{figure*}[tb]
\centerline{\psfig{file=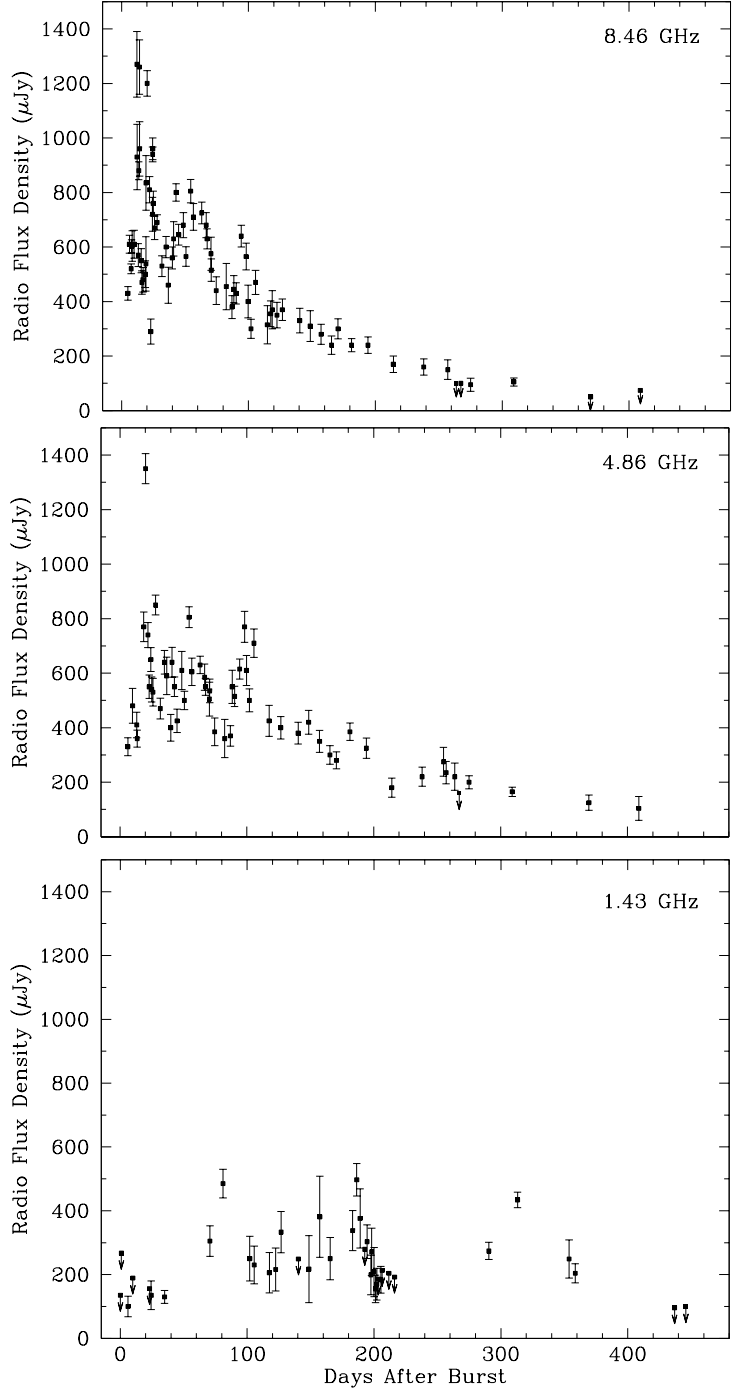,height=20cm}}
\caption[]{The radio light curves of \grb\ at 8.46, 4.86 and 1.43 GHz
  in their entirety. The data are taken from Tables 1 and 2. Upper
  limits are indicated by arrows, with 2-$\sigma$ limits plotted at
  8.46 GHz and 4.86 GHz and 3-$\sigma$ limits plotted at 1.43 GHz.
\label{fig:ltcurve}}
\end{figure*}

\begin{figure*}[tb]
  \centerline{\psfig{file=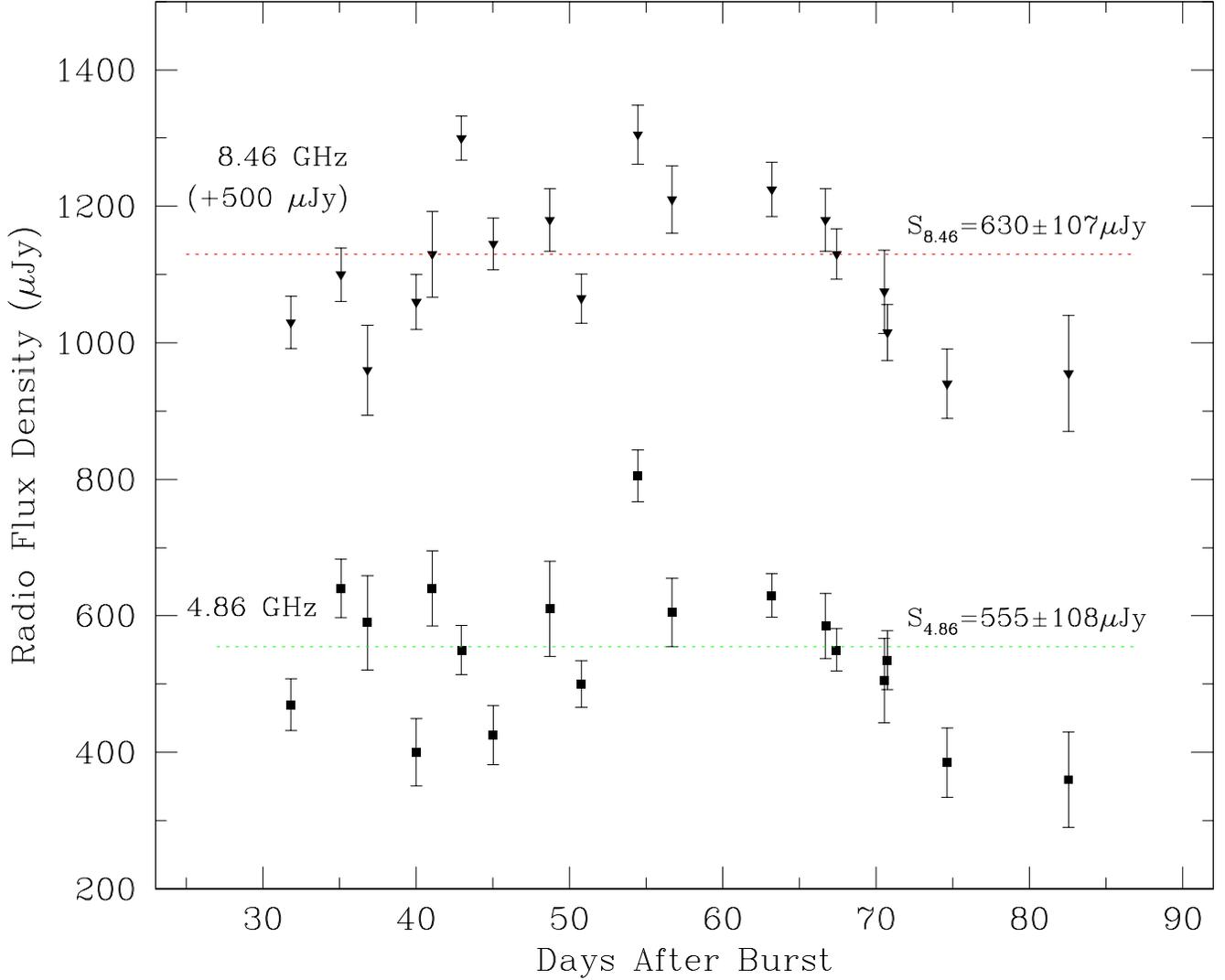,width=18cm}}
\caption[]{Radio light curves of \grb\ at 8.46 and 4.86 GHz in the interval
  between the the quenching of diffractive scintillation and the onset
  of the power-law decay. For ease of viewing, an offset of 500
  $\mu$Jy has been added to the 8.46 GHz data. Flux densities
  indicated on the right hand side are the weighted mean and rms
  scatter of the data in this interval. In the text it is argued that
  the modest, broad-band variations are due to refractive
  scintillation.
\label{fig:cxlight}}
\end{figure*}

\noindent
{\bf Late-Time Light Curves.} In Figure \ref{fig:ltcurve} we show the
entire light curve of all three frequencies at which \grb\ was
monitored by the VLA.  The behavior of the radio light curves in the
first few months has been discussed elsewhere (Frail et al.~1997,
WKF98); Galama et al.~(1998a, b) and  Shepherd et al. (1998) report a
few additional points at low and high frequencies as well.  For the
sake of completeness, we now summarize the early time behavior  here.
As shown in Figure \ref{fig:cxlight}, the erratic flux variations
reported in Frail et al. (1997) do not die away entirely after the
first month but their character changes.  Between 30 d and 85 d the
4.86 GHz and 8.46 GHz light curves have a mean flux density of
555$\pm$108 $\mu$Jy and 630$\pm$107 $\mu$Jy, respectively. The
modulation indices (defined as the ratio of the rms flux variations to
the mean) are modest (15--20\%) and the variations are well correlated
between the two frequencies.  This is very different from the narrow
bandwidth, short time scale fluctuations seen in the first month and
attributed to diffractive interstellar scattering (Frail et al. 1997).
We return to this point in \S\ref{sec:riss}.

The light curve at 1.43 GHz (Table 1 and Figure~\ref{fig:ltcurve}) is
markedly different than the other two.  As noted elsewhere (Frail et
al.~1997, Shepherd et al.~1998, Galama et al.~1998a) the 1.43 GHz flux
density is initially weak (125 $\mu$Jy).  This can best be understood
if the source is optically thick at 1.43 GHz (Katz \& Piran 1997).
Near day 50 the source brightens considerably. Between 50 and 300 days
its average flux density is 300 $\mu$Jy but with considerable scatter.
Most of this scatter likely originates from source confusion,
particularly in the more compact array configurations (see
\S\ref{sec:observations}).

\begin{figure*}[tb]
\centerline{\psfig{file=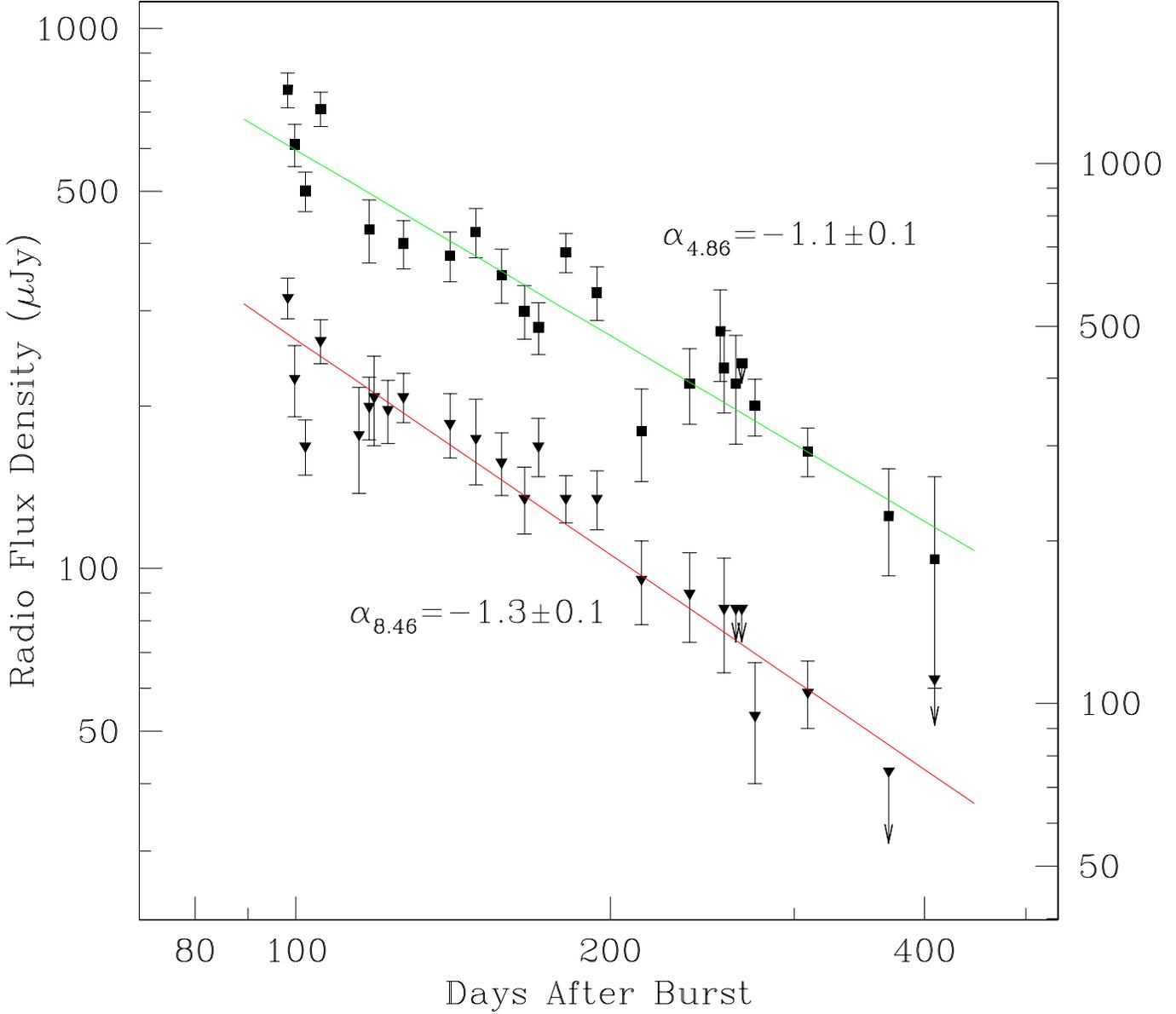,height=20cm}}
\caption[]{Radio light curves of \grb\ at 8.46 and 4.86 GHz in its
  decay phase. The left vertical axis is for 4.86 GHz, while the right
  vertical axis is for 8.46 GHz. Best-fit power law indices $\alpha$
  are indicated by the straight lines.
\label{fig:decay}}
\end{figure*}

Near day 90 the light curves at 8.46 GHz and 4.86 GHz begin a
power-law decay (Figure~\ref{fig:decay}), similar to that which occurred
much earlier at optical and X-ray wavelengths (Pian et al. 1998, Piro
et al. 1998). Power-law fits to the light curves in Figure
\ref{fig:decay} for $t>90$ d give values of
$\alpha_{8.46}=-1.3\pm{0.1}$ and $\alpha_{4.86}=-1.1\pm{0.1}$ (defined
as $f_\nu\propto{t}^\alpha$). The decline at 1.43 GHz is not as well
constrained but we estimate from Figure \ref{fig:ltcurve} that the
decay begin near day 300.

\begin{figure*}[tb]
  \centerline{\psfig{file=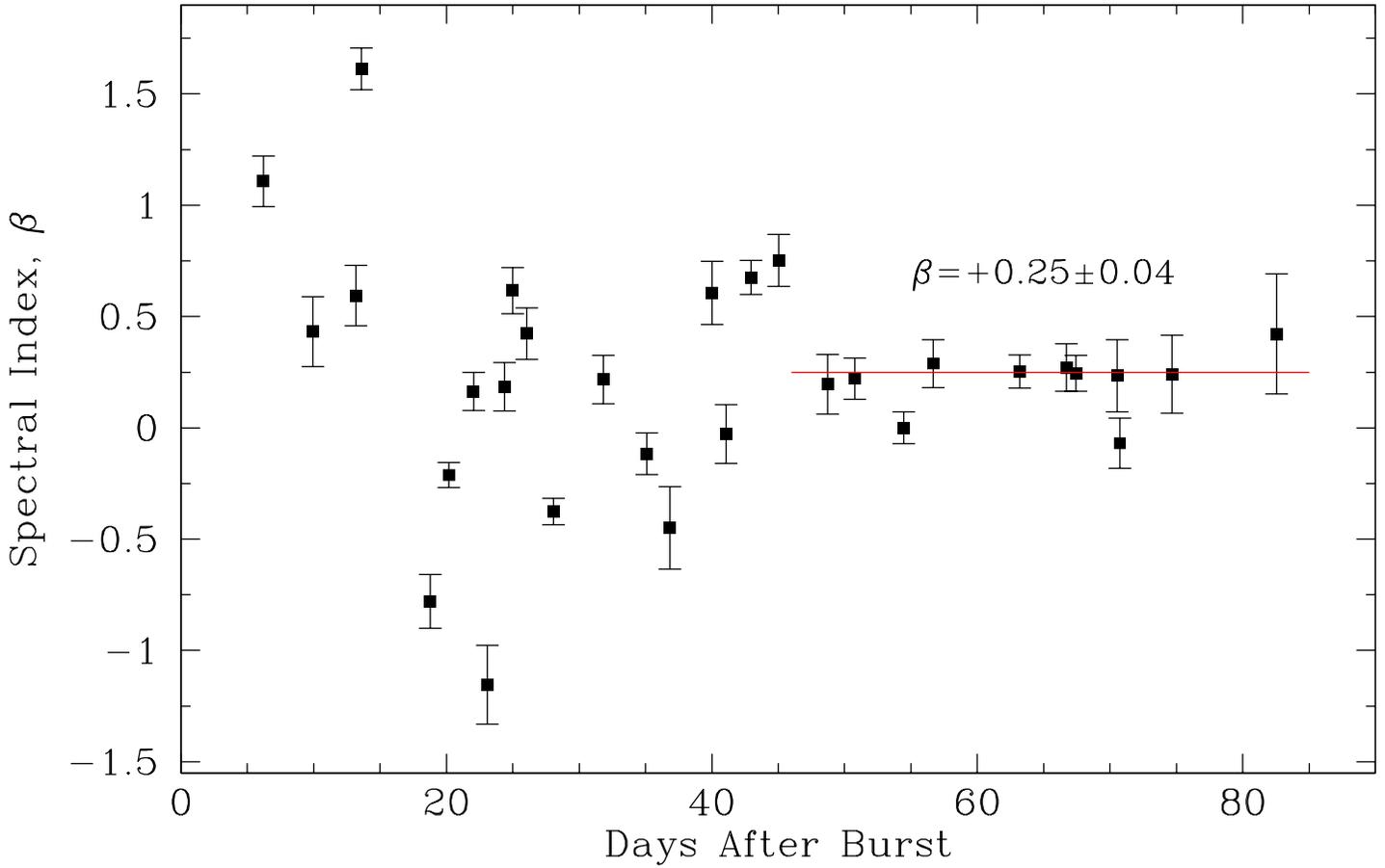,angle=270,width=20cm}}
\caption[]{Spectral index plot between 4.86 GHz and 8.46 GHz showing
  quenching of the narrow-band diffractive scintillation and the
  emergence of the underlying low energy spectrum of the fireball.
  Values of $\beta$ are plotted for all measurements in Table 2 for
  which simultaneous observations were made. The straight line is a
  weighted mean derived over the time range where it is drawn.
\label{fig:cxespec}}
\end{figure*}

\noindent
{\bf Spectral Index.} Many of the features in the high frequency light
curves have analogs in the spectral domain. For example, the narrow
bandwidth ($\delta\nu\sim{4}$ GHz) diffractive scintillation seen by
Frail et al. (1997) is best demonstrated in Figure \ref{fig:cxespec}
by the large swings in the spectral slope $\beta$ (defined as
$f_\nu\propto\nu^\beta$) between 8.46 GHz and 4.86 GHz. The
spectral index appears to settle down by day 35 and we interpret this
to mean that the diffractive scintillation has died down. Indeed,
this is to be expected since only sources smaller than a critical size
can undergo diffractive scintillation. Thus at sufficiently late
times, expansion will ensure that the source enters the refractive 
regime (\S\ref{sec:riss}). Since refractive scintillation is broadband
(see \S\ref{sec:riss}) we see the true radio spectrum of the  afterglow 
past day 35.
Excluding two outlier points,
the mean $\beta$ for 11 measurements between day 40 and day 85 is
+0.25$\pm$0.04 (Figure~\ref{fig:cxespec}). Galama et al.~(1998b)
determined $\beta=0.44\pm{0.07}$ near day 12. Since their estimate is
based on extrapolations of the radio flux density when the source was
scintillating strongly, the value of $\beta$ given here should be a
more accurate one.

In addition, the transition of the light curve to a power-law decay
has a corresponding change in the spectral index plot of Figure
\ref{fig:cxspec}, where $\beta$ undergoes an abrupt drop from positive
to negative values. The weighted mean $\beta$ from day 95 to day 310
is $-0.6\pm$0.3. A joint fit to the spectrum and high frequency light
curves of the form $f_\nu\propto t^{\alpha}\nu^{\beta}$ for $t>110$
days yields more accurate values of $\beta=-0.50\pm0.06$ and
$\alpha=-1.14\pm0.06$ 

\noindent
{\bf Evolution of the Radio Spectrum.} In Figure \ref{fig:allspec} we
follow the evolution of the radio afterglow by plotting all radio
frequencies at different time intervals. These spectral ``snapshots''
show that the spectrum initially rises with frequency, flattens and
then finally exhibits a power-law slope which declines with frequency.

\section{Refractive Scintillation}\label{sec:riss}

In Frail et al. (1997), we attributed the observed strong variability
in the 8.46 GHz and 4.86 GHz bands over the course of the first month,
to diffractive scintillation.  The primary evidence for this inference
is the observed strong, rapid (decorrelation time,
$\delta t < 5$ hr) variability which is
uncorrelated at 8.46 GHz and 4.86 GHz. These two characteristics,
strong and chromatic variability, are
the hallmarks of diffractive scintillation (Rickett 1977).  In
contrast, beyond day 30 the variations are not only milder but also
correlated between the two frequencies (see \S\ref{sec:results}). We
attribute these variations to ``refractive'' scintillation.

Diffractive scintillation is caused by interference between rays
diffracted by small-scale irregularities in the ionized interstellar
medium. The resulting interference is narrow band and highly variable
(with the modulation index approaching unity). Diffractive
scintillation occurs only when the source size is smaller than a
characteristic size, the so-called ``diffractive'' angle. Thus as the
fireball expands we expect the diffractive scintillation to be
quenched. However, as explained by Goodman (1997), focusing and
defocusing of the wavefront by large scale inhomogeneities in the
ionized interstellar medium will also result in scintillation, albeit
at a reduced level on a longer time scale compared to diffractive
variations.  This ``refractive'' scintillation, unlike the earlier
diffractive scintillation, is broad-band, and, most importantly, is
much less sensitive to the size of the source. Thus, not surprisingly,
refractive variability has been seen in many sources, whereas only the
most compact sources -- pulsars and GRB afterglows -- show diffractive
scintillation.

The light curves in Figure \ref{fig:ltcurve} and \ref{fig:cxlight} show
that the expected transition from the diffractive to the refractive
regime did take place.  The flux variations decrease, but do not
vanish, after the first month.  During this period, the modulation
index (defined as the ratio of the rms flux variations to the mean) is
more modest (15--20\%) compared to the first month when the modulation
index was about 80\%.  The change in the nature of the flux variations
is also seen in the spectral index curve (Figure \ref{fig:cxespec}).
Over the first month, the flux variations were uncorrelated between
4.86 GHz and 8.46 GHz (decorrelation bandwidth,
$\nu\leq 3$~GHz), as best demonstrated by
the large swings in the spectral slope $\beta$ whose values varied
randomly between $-1.2$ to +1.6 from one day to the next (Figure
\ref{fig:cxespec}). Starting around day 40, $\beta$ converges to a
constant value.  We identify these slow, broad-band low-amplitude flux
variations as being due to refractive scintillation, with the
expansion of \grb\ taking it through a transition from the diffractive
to refractive regimes.

In principle, an independent estimate of the angular size could be
determined from the refractive scintillation, but our sampling of the
light curve is not sufficient to accurately measure the refractive
decorrelation time scale or the modulation index. 
Nonetheless, as a check for consistency, we now
estimate 
the expected modulation index and refractive time scales and compare
the same to the observations. We
use the angular size of the fireball inferred from the diffractive
scintillations, $\theta\sim 2\, \mu$arcsec at $t=4$ weeks (WKF98),
and extrapolate
this size forward in time as $\theta\propto t^{5/8}$ to obtain a size of
$3\,\mu$arcsec at $t=8$ weeks. From this expected angular size, and
using the formulation of Goodman (1997), we infer the refractive
time scales to be between 10 and 20 hrs and the modulation indices of
order 40\% to 25\% at these frequencies. These estimates are consistent
with our admittedly meager data.

\section{Comparison to Theory: A Relativistic Spherical Model}
\label{sec:Spherical}

As remarked in \S\ref{sec:introduction} the principal motivation for
afterglow observations is to deduce the fundamental parameters of the
explosion.  To this end, we now compare the radio observations with
the expectation of the simplest model, the spherical adiabatic model.
Other authors have carried out similar analyses but based only on
early-time ($\ale 20$~d) multi-wavelength observations (Waxman 1997, Wijers
\&\ Galama 1999, Granot et al. 1999).  They found good agreement with
the basic expectation of spherical adiabatic models.

The first predictive afterglow models (M\'esz\'aros \& Rees 1993,
Paczy\'nski \& Rhoads 1993, Katz 1994, M\'esz\'aros \& Rees 1997)
actually preceded the discovery of the afterglow emission and have
survived with relatively little modification (Sari 1997, Vietri 1997,
Waxman 1997, Wijers, Rees \& M\'esz\'aros 1997). In its most basic
form the cosmological fireball model is a spherical blast wave,
expanding adiabatically into a homogeneous medium of density $n_i$. A
fixed fraction $\epsilon_e$ of the blast wave energy $E_0$ goes
into accelerating a power-law distribution of electrons with energy
slope $p$ (where the number of electrons with Lorentz factor $\gamma$ is 
$dn/d\gamma\propto\gamma^{-p}$ for $\gamma>\gamma_m$).  In the
presence of a magnetic field, which itself is a fixed fraction
$\epsilon_B$ of the energy density of the blast wave, the electrons
emit synchrotron radiation.  The instantaneous synchrotron spectrum
emitted by the electrons consists of four
distinct spectral regimes and is given by Sari, Piran \& Narayan
(1998) for the ``slow cooling'' case when the radiative cooling time
for the majority of electrons with Lorentz factor $\gamma_e\sim\gamma_m$ is
large compared to the dynamical time:

\begin{equation}
f_\nu\propto
   \cases{
     \nu^2 t^{1/2}          & $\nu<\nu_{ab}$,\cr
     \nu^{1/3} t^{1/2}      & $\nu_{ab}<\nu<\nu_m$,\cr
     \nu^{-(p-1)/2} t^{-(3p-3)/4} & $\nu_{m}<\nu<\nu_c$,\cr
     \nu^{-p/2} t^{-(3p-2)/4}     & $\nu>\nu_c$.\cr
         }
\end{equation}

There are three transition frequencies: $\nu_{ab}$, $\nu_m$ and
$\nu_c$.  They correspond to (1) the synchrotron self-absorption
frequency $\nu_{ab}$ below which the source becomes optically thick,
(2) the peak frequency $\nu_m$ where the spectrum peaks at a flux
density of F$_{\nu_m}$ and (3) the cooling frequency $\nu_c$ above
which radiative losses steepen the observed spectrum.  In time, as the
blast wave interacts with the surrounding medium, the shape of the
spectrum is preserved but the peak of the spectrum shifts to lower
frequencies.  The time-dependence of the three transition frequencies
for an adiabatically expanding blast wave is given by
$\nu_{ab}\propto{t}^0$, $\nu_m\propto{t}^{-3/2}$, and
$\nu_c\propto{t}^{-1/2}$ with $\nu_{ab}<\nu_m<\nu_c$ for most of the
the time over which
the afterglow is detectable. In addition, F$_{\nu_m}$ is
expected to remain constant with frequency. Describing the evolution
of the radio flux density as F$_{R}\propto{t}^\alpha\nu^\beta$ for
$\nu_r>\nu_{ab}$ (i.e. the optically thin regime) we expect to see a
slow rise to F$_{\nu_m}$ as $t^{1/2}$ for $\beta$=+1/3, followed by a
power-law decline going as $t^\alpha$, where $\alpha=3\beta/2$ and
$\beta=-(p-1)/2$ when $\nu_m\leq\nu_r$; and $\alpha$ steepening a
further $-0.25$ when $\nu_c\leq\nu$.

The observations presented in \S\ref{sec:results} show substantially
all of the gross spectral and temporal behavior expected from this
basic model. In detail we find the following agreements with the
model.

\begin{figure*}[tb]
  \centerline{\psfig{file=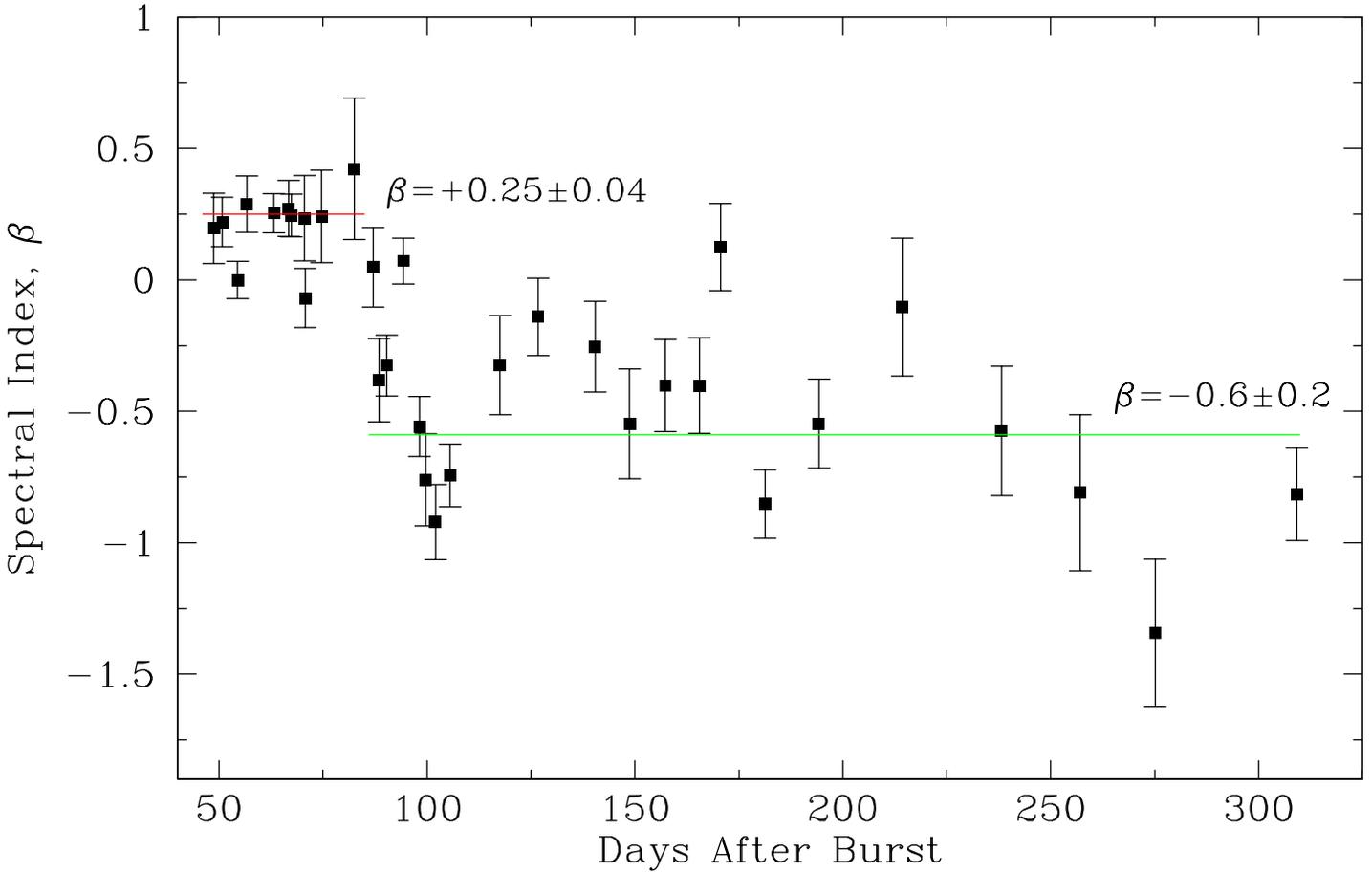,angle=270,width=20cm}}
\caption[]{Spectral index plot between 4.86 GHz and 8.46 GHz.
  Values of $\beta$ are plotted for all measurements in Table 2 for
  which simultaneous observations were made. The straight lines are
  weighted means derived over the time range where they are drawn.
\label{fig:cxspec}}
\end{figure*}

\noindent{\bf Low Energy Spectral Index.}
At early times synchrotron self-absorption is clearly detected
(Figure~\ref{fig:allspec} and Shepherd et al.~1998) with a synchrotron
self-absorption frequency $\nu_{ab}$ near 3 GHz (Granot et al.~1999).
After day 35 when the diffractive interstellar scintillation between
4.86 GHz and 8.46 GHz has quenched, the low energy regime of the
synchrotron spectrum (i.e. $\nu_{ab}<\nu<\nu_m$) is revealed. The
measured slope of +0.25$\pm$0.04 (Figure~\ref{fig:cxspec}) is close to
the predicted asymptotic value of 1/3. This slope is a generic
prediction of relativistic shocks which accelerate electrons with a
power-law distribution of energies and have sharp cutoffs at low
energies (Katz 1994).

\noindent{\bf Temporal Decay: Indices and Time scales.} The decay in
the 8.46 GHz and 4.86 GHz light curves (Figure  \ref{fig:decay}) begins
near day 90 and can be most easily seen in the spectral index plot of
Figure \ref{fig:cxspec}, where the spectral slope undergoes an abrupt
change from positive to negative values corresponding to the regime
$\nu\geq\nu_m$. The corresponding light curves have a power-law
decay with a temporal slope near $-1.2$, comparable to that inferred
from optical and X-ray observations (see Table 4).
 
As expected, the decay of the light curve for the lower frequency of
1.43 GHz begins much later (Figure \ref{fig:ltcurve}). We can compare
the time scale for the onset of the decay with the predicted $t^{-3/2}$
dependence for $\nu_m$. In Table 3 we give estimates of $t_m$ between
1 and 100 GHz.  The $t_m$ from the 86 GHz data of Bremer et al. (1998)
would predict that the onset of the power-law decay would start
between 50--70 d at 8.46 GHz and between 70--95 d at 4.86 GHz.
Similarly, from the 8.46 GHz and 4.86 GHz data we would predict a
decay at 1.43 GHz between 200--300 d after the burst. Given the
uncertainties in the estimates of $t_m$ this agreement is surprisingly
good.

\noindent{\bf Lack of cooling break.} There is no evidence for 
the existence of an additional cooling break in the light curves  or
spectra of Figures \ref{fig:decay} and \ref{fig:cxspec}.
In contrast, Galama et al. (1998b,c)
have identified passage of the cooling break at higher frequencies,
specifically in the optical band around day 1.4. Accepting this result,
the expected epoch of cooling break in the 8.46 GHz band is
$t_c(O)(\nu_O/\nu_R)^{2}$, which is considerably larger than one
year; here $\nu_O$ is the frequency in the optical band, $\sim 5\times
10^{14}$ Hz, $\nu_R$ is the typical radio frequency 
and $t_c(O)\sim 1.4$ d.  
Thus, the absence of any clear cooling break in our
radio light curves is not surprising.
 
\section{Deviations From the Spherical Relativistic Model}
\label{sec:Deviations}

In the previous section we found that the spherical adiabatic fireball
model provided a satisfactory explanation for the observations at a
gross level. However, on closer inspection there are several important
differences between the radio afterglow of \grb\ and the model
predictions. Indeed, on theoretical grounds we expect deviations from
the spherical relativistic model due to a transition from relativistic
to sub-relativistic regimes and possibly jet geometry (WKF98). We
summarize the observed deviations below and offer explanations for them
in \S\ref{sec:sub-Relativistic}.

\noindent{\bf Deviation in the Light Curves.} As already noted in
WKF98, the spherical model which provided an excellent fit to the
early time (less than 3 weeks) multi-wavelength data (X-ray, optical
and radio), predicted radio emission that was well above the observed
value for $t>25$ d; see Figure 1 of WKF98.

\noindent{\bf Non-constancy of Peak Flux.} Another important
prediction of the spherical, adiabatic model is the constancy of the
peak flux density F$_{\nu_m}$ with time. While the peak frequency
$\nu_m$ evolves as $t^{-3/2}$, the value of F$_{\nu_m}$ is expected to
remain constant as $\nu_m$ sweeps through the X-ray, optical and radio
bands.  This assumption underlies all attempts to fit ``snap-shot''
broad-band spectral distribution of GRB afterglows (e.g. Galama et al.
1998b, Wijers \& Galama 1999) and the subsequent derivation of the
shock parameters. In Table 3 there is a very clear trend for
successive peaks in F$_{\nu_m}$ to reach lower peak flux densities as
$\nu_m$ moves towards lower frequencies. A least squares fit between 1
and 90 GHz gives F$_{\nu_m}\propto\nu_m^{0.40\pm{0.04}}$.

\begin{figure*}[tb]
\centerline{\psfig{file=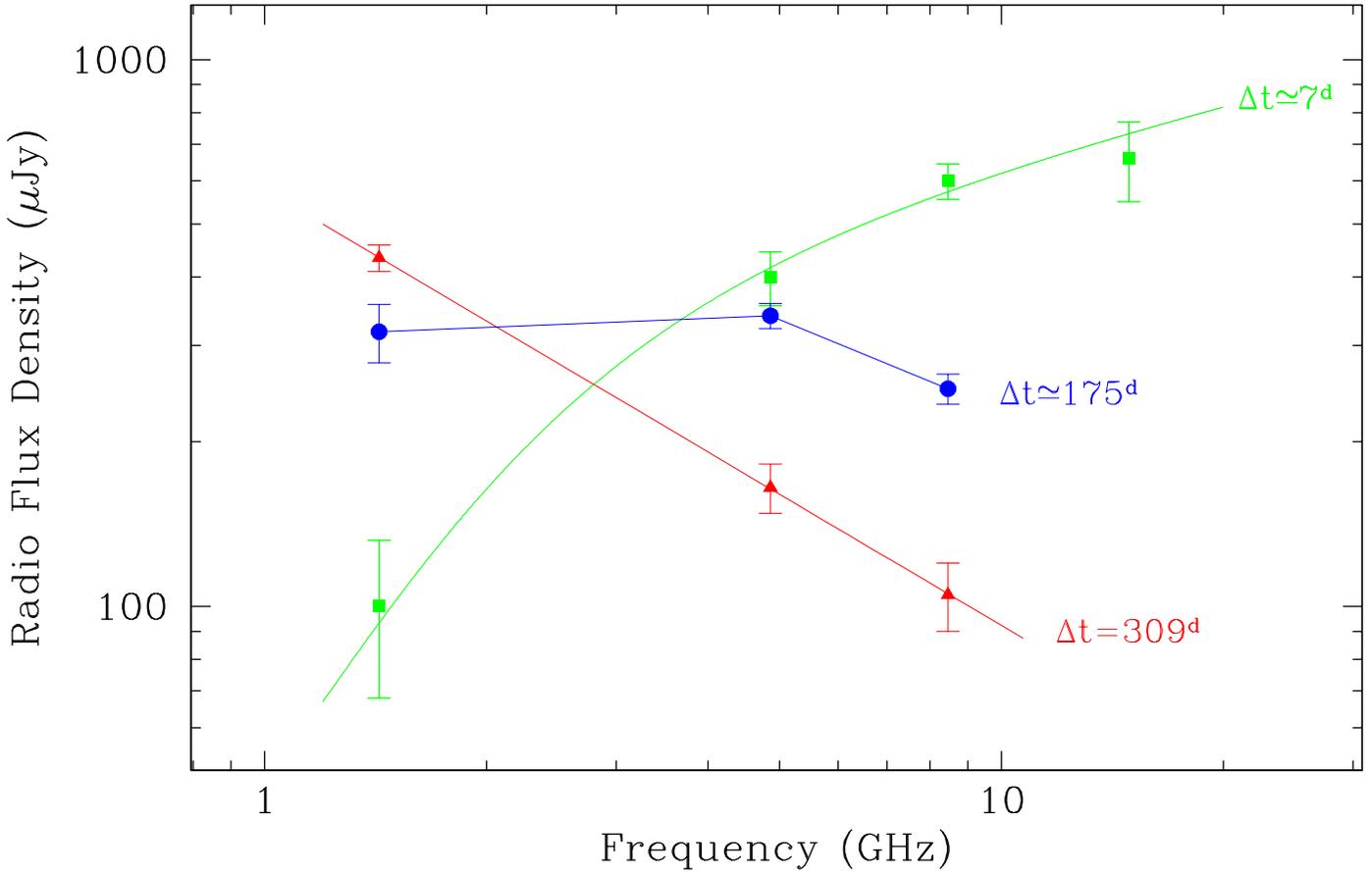,angle=270,width=20cm}}
\caption[]{Spectral ``snapshots'' showing the
  evolution of the \grb\ in the radio band. The data points (solid
  squares) from day 7 are taken from Shepherd et al. (1998) and fit to
  the parameters derived by Granot et al. (1998). The data from near
  day 175 are averages taken around this time. The data at day 309 was
  taken just prior to the decay of the 1.43 GHz light curve and ar fit
  to a power law $\nu^{-0.8\pm{0.1}}$.
\label{fig:allspec}}
\end{figure*}

\noindent{\bf Non-constancy of Absorption Frequency.}
In the simple adiabatic fireball model, provided that the fundamental
shock parameters (E$_0$, $n_i$, $\epsilon_e$ and $\epsilon_B$) do
not change, $\nu_{ab}$ is expected to remain constant until such time
as $\nu_m\leq\nu_{ab}$ (Waxman 1997, Granot et al. 1999). Near day 50
the 1.43 GHz flux density rises from $\sim$125 $\mu$Jy to $\sim$300
$\mu$Jy (see Figure  \ref{fig:ltcurve}). Such behavior is expected for
an expanding optically thick source (see Frail et al.~1997) simply
because of the increase in the radiative surface area.  However, the
spectra in Figure \ref{fig:allspec} suggest a different
interpretation. The initial attenuation at $t=7$ d is well described
by synchrotron self-absorption with unity optical depth at
$\nu_{ab}$=3 GHz (Shepherd et al. 1998, Granot et al.  1999). In time,
the spectra in Figure \ref{fig:allspec} clearly show an evolution from
an optically thick to an optically thin source, with $\nu_{ab}$
shifting to lower frequencies.

\noindent{\bf Relation between $\alpha$ and $\beta$.} The final and
perhaps the most significant deviation from the model concerns the
measured values for $\alpha$ and $\beta$ (where
$f_\nu\propto{t}^\alpha\nu^\beta$) of $\alpha=-1.14\pm0.06$ and
$\beta=-0.50\pm0.06$ from the decay of the radio light curve.  In
Table 4 these values are compared to measurements made at optical
wavelengths at different times in the evolution of the light curve.
The temporal slope of the optical light steepens from its initial
value of $\alpha=-0.9$ (Djorgovski et al.~1997) to $-1.2$ (e.g.
Zharikov et al.  1998, Garcia et al. 1998, Castro-Tirado et al.  1998,
Bloom et al. 1998). Similarly, the spectral slope undergoes an abrupt
transition at optical wavelengths between day 1 and day 2 from
$\beta=-0.6$ (Djorgovski et al. 1997) to $\beta=-1.1$ (Galama et al.
1998c, Sokolov et al. 1998). Galama et al (1998b) have made a strong
case that this spectral and temporal break corresponded to the passage
of the cooling break $\nu_c$ through the optical passbands at roughly
1.4 d after the burst. Prior to the passage of $\nu_c$ the
relativistic model predicts $\alpha=3\beta/2$, followed by a
steepening (by 1/2) of $\beta$ giving $\alpha=3\beta/2$ + 1/2. The
measured $\alpha$ and $\beta$ on either side of $\nu_c$ values are
consistent with these relationships. Bolstering this claim, Galama et
al.~(1998a) derived two independent measures of the electron energy
power-law index ($p\simeq{2.2}$). In contrast, in the framework
of the relativistic adiabatic model, no single value of $p$
can explain the values of $\alpha$ and $\beta$ that we obtained from
our radio observations.
The radio $\alpha$
is identical to the late-time optical value but the radio $\beta$ is
more similar to the early optical estimates before $\nu_c$ moved
through the optical band.

\section{Jets and the Transition to the Sub-relativistic Regime }
\label{sec:sub-Relativistic}

The relativistic spherical model is the simplest afterglow model and
as such has received considerable attention from observers.  Apart
from the assumptions of constant $\epsilon_e$, $\epsilon_B$ and
acceleration to a power law spectrum, the main assumptions of the
model are spherical geometry and relativistic outflow. Hints of
deviations from the expectations of the spherical model were already
apparent in the WKF98 paper (see \ref{sec:introduction}).  

In this section, we first qualitatively expand on these two ideas and
find that the four problems identified in the previous section can be
satisfactorily explained by a model incorporating jets. An inevitable
consequence of a jet geometry is early (compared to spherical
geometry) transition to the non-relativistic regime.
We develop and present in Appendix \ref{sec:NRModel}
a formulation describing non-relativistic fireballs, and then 
apply this formulation to the 
light curves (Figure \ref{fig:ltcurve}) and extract
the basic parameters of the explosion in \S\ref{sec:Calorimetry}.

\noindent{\bf Deficit of Flux and Lack of Constancy of $F_{\nu_m}$: Jets.}
A jet with an opening angle of $\theta_j$ behaves as if
it were a conical section of a spherical fireball, and cannot be
distinguished by a distant observer from a spherical fireball, as long
as the expansion Lorentz factor is larger than the inverse of the
opening angle, $\gamma>\gamma_j\sim \theta_j^{-1}$. Once the jet
decelerates to $\gamma<\gamma_j$, it expands sideways, leading to a
decrease in blast wave energy per unit solid angle, which results in a
deviation from spherical fireball behavior. The decrease in energy per
unit solid angle leads to a decrease with time of $F_{\nu_m}$, and
suppresses the $t^{1/2}$ flux increase expected in the spherical model
at frequencies $\nu<\nu_m$. Thus a jet geometry provides a reasonable
explanation for the flattening of the radio light curves and the lack
of constancy of $F_{\nu_m}$ (\S\ref{sec:Deviations}).

We set $t_j$, the epoch when the jet becomes apparent to the observer,
as 25 d (see Figure 1 of WKF98; also discussion in
\S\ref{sec:Deviations}).  The Lorentz factor corresponding to this
epoch is $\gamma_j\sim 2$, or equivalent to a jet opening angle
$\theta_j\sim 30^\circ$.  It was shown in WKF98 that on a time scale of
$\gamma_j^2 t_j$, where $t_j$ is the time at which the jet decelerates
to $\gamma=\gamma_j$, the fireball approaches spherical,
sub-relativistic expansion. A simple argument
to understand this is as follows. The
time at which we see deviation from spherical symmetry is
$t_j\sim{r_j}/(2\gamma_j^2 c)$, where $r_j$ is the radius at which
$\gamma=\gamma_j\simeq \theta_j^{-1}$.  The transition to spherical
symmetry occurs over time scale $t_s\sim{r_j}/c$, i.e.  $t_s\sim\gamma_j^2
t_j$. In addition, since the mass enclosed within a sphere of radius
$r_j$, $M_j=4\pi r_j^3 n_i m_p/3$, is comparable to the total jet energy 
$E_j\simeq E_{\rm sph.}/2\gamma_j^2$, where 
$E_{\rm sph.}=16\pi r_j^3 \gamma_j ^2 n m_p c^2/17$ is the energy inferred
under the assumption of spherical symmetry, the fireball becomes 
sub-relativistic as it approaches spherical symmetry.
Using our $t_j\sim$25~d and $\gamma_j\sim$2 we expect the fireball to 
approach spherical sub-relativistic expansion on a time scale 
of $t_s\sim$100\,d.

\noindent{\bf Relation between $\alpha$
and $\beta$: Sub-relativistic
Fireball}. As discussed in Appendix \ref{sec:NRModel}, in the sub-relativistic
regime, the afterglow emission scales as $f_\nu\propto
t^{\alpha}\nu^{\beta}$ with $\beta=(p-1)/2$ and
$\alpha=3\beta+3/5$ (compared to
$\alpha=3\beta/2$ for relativistic expansion).  The behavior of the
radio flux observed at $t>100$~d is consistent with this expectation
since past day 100 (approximately the transition to sub-relativistic
regime) we obtain $\alpha=-1.14\pm 0.06$ and $\beta=-0.50\pm 0.06$
(\S\ref{sec:results}).  If the afterglow was in the sub-relativistic
regime, then we expect the following closure relation,
$N=\alpha-3\beta-3/5\equiv{0}$, whereas in the relativistic regime,
$R=\alpha-3\beta/2\equiv{0}$. We find $N=-0.24\pm 0.19$ whereas
$R=-0.39\pm 0.11$. Thus we can reject the relativistic model at the
$\sim$3-$\sigma$ level.

More importantly, a single value of $p$=2.2, explains, in a
self-consistent way, the X-ray (Piro et al.~1998), optical (Galama et
al. 1998b) and radio light (this paper) afterglow observations over
several decades in time through radiative transitions (adiabatic to
cooling) and dynamical transitions (relativistic to sub-relativistic).
%In contrast, if we adopt the relativistic model,
%from the observed value of $\alpha$ we derive $p=2.5$ which is an
%unusual steep value of $p$ and is inconsistent with determinations
%earlier observations.

\noindent{\bf Non-Constancy of $\nu_{ab}$.}
The non-constancy of $\nu_{ab}$ can be explained as follows.
There are two consequences in the sub-relativistic
model when $\nu_m$ falls below $\nu_{ab}$
occurs (see WKF98 and Appendix \ref{sec:NRModel}). First, $\nu_{ab}$
is no longer constant but evolves as $\propto t^{-2/3}$. Second, the
peak in the spectrum is no longer at $\nu_m$ but is at $\nu_{ab}$. The
peak flux then evolves as $t^{-17/30}$. By itself, the observed
decrease in $\nu_{ab}$ is consistent with the expectations of the
sub-relativistic model. However, since one expects a similar temporal
evolution of $\nu_{ab}$ for $\nu_m\leq\nu_{ab}$ in the relativistic
model, it should not be taken as strong evidence in favor of the
transition to sub-relativistic expansion.

The above conclusions, the manifestation of a jet at $t_j\sim 25$ d
and the sub-relativistic transition at about epoch 100~d have been
arrived at using extensive radio data presented in this paper. 
Independent support for
this can be obtained from the optical data. As noted by Rhoads (1999)
the optical data cannot be fitted by a single power law decay
functional form. Rhoads derives a minimum beaming angle of
$\theta_j=30^\circ$, a value in complete agreement with our own. 
The principal strength of the radio result is the photometric uniformity
of the data -- the data were obtained with one instrument, the VLA, in a 
standard mode. In contrast, the optical data come from diverse sources
and vexing cross-calibration issues still remain.  

In our model, we would predict that the optical flux would steepen
beyond $t_j$.  A sharp break, such as the one seen in the light curve
of GRB\thinspace{990510} (e.g.  Harrison et al. 1999), is not expected
in the case of \grb, given the wide opening angle.
Numerical and analytical studies which track the
dynamical evolution of the beamed ejecta (Moderski, Sikora, \& Bulik
1999, Wei \& Li 1999) find that unless the opening angle is very small
or that lateral expansion is unimportant, a smooth and gradual
transition is expected.  In addition, past day 100, our model would
predict a further, slight steepening [cf. Eq. (\ref{eq:f_nuc})]
of the optical flux as the shock
makes the transition from the relativistic to sub-relativistic
regimes.  The combination of these two factors is the likely origin for
the apparent disappearance of the afterglow at $t\simeq 450$ d in a
recent HST image (Fruchter et al. 1999).

\section{Calorimetry of the Explosion}
\label{sec:Calorimetry}

In Appendix \ref{sec:NRModel}, we present a rigorous formulation of a 
sub-relativistic fireball model.  In WKF98
we considered sub-relativistic fireball models in the asymptotic limit
of $\nu_{ab}$ being very different from $\nu_m$.  However, for the
data presented here, $\nu_m$ and $\nu_{ab}$ are comparable, and the
model presented in Appendix \ref{sec:NRModel} does not rely on asymptotic 
limits but is an exact model.

\begin{figure*}[tb]
\centerline{\psfig{file=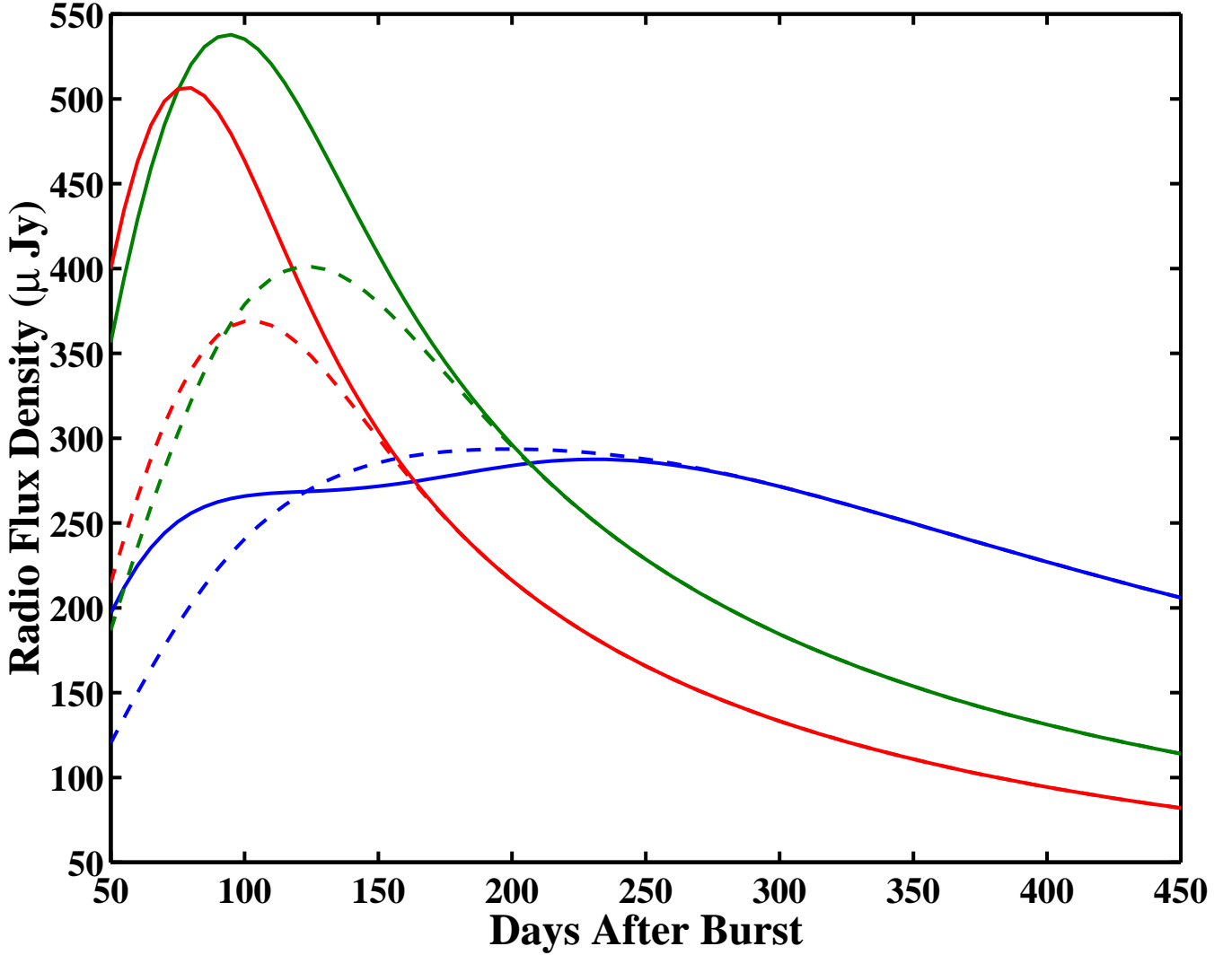,height=15cm}}
\caption[]{Model light curves for $a=4.1\times10^{-51}
  {\rm\ g\ s}^{1/2}$, $b=5.3\times10^{29}{\rm\ s}^{-3.1}$ and
  $\nu_0/(1+z)=2.3$~GHz (solid curve), $\nu_0/(1+z)=5.2$~GHz (dashed
  curve). Both models reproduce the normalization and power law time
  dependence of the 4.86 and 8.46~GHz fluxes (Figure \ref{fig:decay}), 
  as well as the early
  time suppression of flux at 1.43~GHz. In both models the optical
  depth at 1.43~GHz initially increases, and then rapidly decreases as
  $\nu_{ab}$ drops below $\nu_m$ (cf. Eqs. \ref{eq:nu_A1},
  \ref{eq:f_p2}), allowing both models to correctly reproduce the
  approximately constant flux at 1.43~GHz at $t=100$~d to $t=300$~d
  (Figure \ref{fig:ltcurve}).
  Both models also correctly predict the decrease of 1.43~GHz flux at
  $t>300$~d (Figure \ref{fig:ltcurve}).
\label{fig:ewmodel}}
\end{figure*}

The radio flux and spectrum produced by the sub-relativistic fireball 
are determined by three parameters,
$a$, $b$ and $\nu_0$ 
(see Appendix \ref{sec:NRModel}). $\nu_0$ is $(1+z)\nu_m$ at
epoch $t_0$. Here we set $t_0=100\,$ d and note that the redshift of the
host galaxy of \grb\ is $z=0.835$ (Bloom et al. 1998). The $t>$100~d
radio data allow an accurate ($\pm 10\%$) determination 
of the parameters $a$ and $b$, which determine the high frequency flux
normalization and the self-absorption suppression at 1.43~GHz:
$a=3.7\times10^{-51}{\rm\ g\ s}^{1/2}$, $b=5.3\times10^{29}{\rm\ 
  s}^{-3.1}$. The spectral change at $t\simeq 100$~d (Figure
\ref{fig:cxspec}) implies that the peak frequency $\nu_m$ decreases
through the frequency range of 8.46~GHz to 4.86~GHz at $t\simeq
100$~d, i.e. that $\nu_0/(1+z)\simeq 5$~GHz.  However, it should be
noted that since in our model the fireball at time $t=100$~d is in
transition to sub-relativistic spherical expansion (see below),
significant deviation from the sub-relativistic model prediction is
possible at $t=100$~d. Thus, the determination of $\nu_0$ is only
approximate (see Figure \ref{fig:ewmodel}).  Our conclusions remain
unchanged, however, even with a factor of a few change in $\nu_0$ (see
below).

The parameters $a$, $b$, and $\nu_0$ are determined (see 
appendix \ref{sec:NRModel})
by the four parameters, which completely specify a sub-relativistic fireball
model: the strength $B_0$ of the magnetic field at
$t=t_0$, the minimum Lorentz factor 
$\gamma_0$ of the radiating
electrons at $t=t_0$, the (time independent)
number density $n$ of the shocked plasma, 
and the shock radius $r_0$ at $t=t_0$. 
Using the three constraints provided by the determination 
of $a$, $b$ and $\nu_0$ we first derive
$B_0$, $\gamma_0$, and $n$ as
functions of $r_0$. We then constrain $r_0$ by requiring that the
derived minimum energy of the radiating electrons and magnetic field does
not exceed the thermal energy of the sub-relativistic fireball, given by 
the Sedov-von Neumann-Taylor formulation.

Using equation~(\ref{eq:a}) we find
\begin{equation}
B_0=5.2\left({r_{18}\over d_{\star}}\right)^4 a_\star^{-2} {\rm\ G}.
\label{eq:B}
\end{equation}
We use the definitions $r_{18}=(r_0/10^{18}~{\rm cm})$,
$a=4.1a_\star\times10^{-51}{\rm\ g\ s}^{1/2}$,
$b=5.3b_\star\times10^{29}{\rm\ s}^{-3.1}$,
$\nu_0/(1+z)=5.2\nu_\star$~GHz, and
$d=0.94d_\star\times10^{28}$~cm. As demonstrated above,
observations require $a_\star=b_\star=1$, and $\nu_\star\approx1$.
$d_\star=1$ for a flat universe with
zero cosmological constant and $H_0=70\,{\rm km\, s}^{-1}\,{\rm Mpc}^{-1}$.
The minimum Lorentz factor is obtained from
equation~(\ref{eq:nu0}),
\begin{equation}
\gamma_0=25.7 \nu_\star^{1/2} \left({r_{18}\over d_{\star}}\right)^{-2}
a_\star.
\label{eq:gamma_0}
\end{equation}
Finally, substitution of the above results into equation~(\ref{eq:b})
yields,
\begin{equation}
n=0.005\eta_1\nu_\star^{-0.6} b_\star a_\star^3 r_{18}^{-7}
d_{\star}^6~{\rm cm^{-3}},
\label{eq:n}
\end{equation}
where $\eta=10\eta_1$; 
here $r/\eta$ is the thickness of
the shell of the fireball.

The ratio of electron to magnetic field energy densities is
\begin{equation}
{E_e\over E_B}=\frac{p-1}{p-2}{n\gamma_0 m_e c^2\over B^2/8\pi}
=5.4\times10^{-7}\eta_1
\nu_\star^{-0.1} b_\star a_\star^8 r_{18}^{-17} d_{\star}^{16}.
\label{eq:e_B}
\end{equation}
The minimum combined electron and magnetic field energy,
$E_{min}=1.92E_{ep}$ is obtained for
a radius $r=r_{min}=0.96r_{ep}$, where the equipartition radius $r_{ep}$
is the radius for which $E_e=E_B=E_{ep}$. Using the above equations
we obtain
\begin{equation}
r_{ep}=4.3\times10^{17}\eta_1^{1/17}
\nu_\star^{-1/170} b_\star^{1/17} a_\star^{8/17} d_{\star}^{16/17}
{\rm\ cm},
\label{eq:r_ep}
\end{equation}
and
\begin{equation}
E_{min}=2.3\times10^{50}\eta_1^{-6/17}
\nu_\star^{-11/170} b_\star^{11/17} a_\star^{20/17} d_{\star}^{40/17}
{\rm\ erg}.
\label{eq:E_ep}
\end{equation}

During sub-relativistic expansion, the fireball hydrodynamics is described 
by the self-similar Sedov-von Neumann-Taylor solution. 
We now compare the equipartition energy with the energy of a Sedov-von
Neumann-Taylor
shock which reaches radius $r_0$ at time $t_0$. This energy
is given through the relation
\begin{equation}
r_0=\xi(\hat\gamma)\left({E_{ST}\over n_i m_p}\right)^{1/5}
\left({t_0\over1+z}\right)^{2/5}.
\end{equation}
Here $n_i$ is the ambient medium density,
$\hat\gamma$ is the adiabatic index of the
gas and $\xi$ is close to unity for $\hat\gamma$ in the range of
$4/3$ (relativistic fluid) and $5/3$ (non relativistic fluid). In
what follows we use $\hat\gamma=13/9$, appropriate for pressure
equilibrium between relativistic electrons and non relativistic protons,
for which $\xi(\hat\gamma=4/3)=1.05$. Using the above equations we have
\begin{equation}
E_{ST}=4.4\times10^{50}\eta_1^{-2/17}\nu_\star^{-0.59}
b_\star^{15/17} a_\star^{35/17} d_{\star}^{70/17}
\left({r_0\over r_{ep}}\right)^{-2}{\rm\ erg}.
\label{eq:E_ST}
\end{equation}
At $r=r_{ep}$ we find that the Sedov-Taylor energy is about twice the
minimum electron and magnetic field energy required to account for the
radio emission, $E_{ST}(r_0=r_{ep})\simeq 2E_{min}$.  

The thermal energy in a Sedov-Taylor solution constitutes approximately
half the total energy. The energy in the shocked electrons
and in magnetic field clearly cannot exceed the thermal energy.
Thus, we obtain an additional constraint, $E_{min} \leq E_{ST}/2$.
As can be seen from the above equations this constraint can only
be satisfied with $r_0\sim r_{ep}$.
For larger or smaller $r_0$, $E_e+E_B$ would exceed the shock thermal
energy, as $E_{ST}\propto (r_0/r_{ep})^{-2}$, while
$E_e\propto(r_0/r_{ep})^{-6}$ and $E_B\propto(r_0/r_{ep})^{11}$.

Thus, although radio observations provide only three constraints
on the four model parameters, a self-consistent solution is
obtained only for
\begin{equation}
r_0\simeq r_{ep}=4.3\times10^{17}\eta_1^{1/17}
\nu_\star^{-1/170} b_\star^{1/17} a_\star^{8/17} d_{\star}^{16/17}
{\rm\ cm}.
\label{eq:r_0}
\end{equation}
This implies that electrons and magnetic fields are close to equipartition,
that the fireball energy is
\begin{equation}
E\simeq E_{ST}(r_0=r_{ep})
=4.4\times10^{50}\eta_1^{-2/17}\nu_\star^{-0.59}
b_\star^{15/17} a_\star^{35/17}d_{\star}^{70/17}{\rm\ erg},
\label{eq:E}
\end{equation}
and that the ambient density is
\begin{equation}
n_i\simeq\frac{3}{\eta}n(r_0=r_{ep})=
0.53\eta_1^{10/17}\nu_\star^{-0.56} b_\star^{10/17} a_\star^{-5/17}
d_{\star}^{-10/17} {\rm\ cm}^{-3}.
\label{eq:n_i}
\end{equation}

Interstellar scintillations provide a direct and independent
confirmation of the derived radius.  As noted in WKF98,
the observations of scintillation yield a fireball radius, 
$r\simeq3\times10^{17}$ cm
at $t=25$~d which compares favorably  with that stated
in Equation \ref{eq:r_0}.
Furthermore, the shock radius 
derived in Equation (\ref{eq:r_0}) implies a shock velocity
\begin{equation}
\beta_{ST}\equiv{2r\over 5ct/(1+z)}=0.6
\left({t\over300{\rm\ d}}\right)^{-3/5}.
\label{beta_ST}
\end{equation}
Thus, our model predicts relativistic to sub-relativistic transition
at $t\sim100$~d --- as was indeed inferred from the $t<100$ d data.
We also note that, assuming jet expansion at $t\simeq 25$~d,
the inferred fireball
energy is $\ale 10^{51}$ erg, rather than the isotropic $\sim 10^{52}$
erg, and in excellent agreement with the value obtained using the
$t>$100~d data.

\section{An Alternate Model}\label{sec:Alternate}

Based on these same radio data, Chevalier \& Li (1999a) have proposed
a model in which the afterglow is produced by the expansion of a shock
into a circumburst medium, shaped by the mass loss history of the
massive progenitor star.  In this ``wind'' model, one expects the
ambient density to be $\propto r^{-2}$ whereas the density is expected
to be constant for an explosion in the typical interstellar medium.
There is increasing support for such models, both in direct evidence
linking supernovae with gamma-ray bursts (Galama et al. 1998d,
Kulkarni et al. 1998), and indirect evidence in the optical and radio
light curves of the afterglows (e.g. Bloom et al.~1999, Reichart 1999,
Galama et al. 1999, Chevalier \& Li 1999b, Frail et al.~1999).  We
find, however, that a wind model is inconsistent with \grb\ data.

Let us first consider the temporal dependence of the optical and radio
flux. Of particular importance, is the location and subsequent
evolution of the cooling frequency at early times. For an electron
power-law index $p=2.2$, implied by early afterglow observations
(Galama {\it et al.} 1998a,b) and used by Chevalier \& Li, the flux at
frequencies above the peak frequency makes a transition in the wind
model from a $t^{-1.15}$ decline to $t^{-1.4}$, as the cooling
frequency $\nu_c$ passes through the band (with
$\nu_c\propto{t^{1/2}}$). In contrast, for the same value of $p$, in
the constant density model, the cooling frequency decreases in
time (as $\nu_c\propto{t^{-1/2}}$) and the flux at frequencies above
the peak frequency makes a transition in the wind model from a
$t^{-0.9}$ decline to $t^{-1.15}$.

Early afterglow observations (Galama et al. 1998b) yield two
independent estimates of $\nu_c$, the first at $4.5\times10^{14}$~Hz
at $t=1.4$~d and another at $1.6\times10^{14}$~Hz at $t=12.1$~d.
Chevalier \& Li (1999a) have questioned the identification of $\nu_c$
at $t=1.4$~d, since the optical light curve prior to this time
deviates significantly from light curves of afterglow models which
account for $t>1.5$~d data. However, the existence of $\nu_c$ dropping
through the optical band on time scale of days following the burst can
also be inferred from infra-red observations (Pian et al. 1998, Chary
et al. 1998) at $t>3$~d. The Galama et al. observations imply that
$\nu_c$ {\it decreased} in frequency from 1.4 days to 12.1 days, in
clear contradiction with expectations for the wind model. Furthermore,
since these observations imply that $\nu_c$ had passed through the
optical bands in the first few weeks, in the wind model both the
optical flux (at $t>10$~d) and the radio flux (at $t>100$~d) should
decline as $t^{-1.4}$. Such behavior is in contradiction with
observations, which show a $t^{-1.15}$ flux decline at both optical
and radio frequencies (Note that the optical flux decline in the wind
model can be modified to follow a $t^{-1.15}$ power law by modifying
model parameters to give $\nu_c\ll4\times10^{14}(t/10d)^{1/2}$~Hz, so
as to have $\nu_c$ below the optical band. However, such choice would
be inconsistent with the value of $\nu_c$ inferred from observations).
Moreover, for the energy and density required to account for
$t\sim10$~d observations, $E\sim10^{52}$~erg and $n\sim1{\rm
  cm}^{-3}$, the wind Lorentz factor at $t=100$~d is $\gamma\simeq1.2$
(the Lorentz factor is only weakly dependent on $E$ and $n$, and for
the parameter choice of Chevalier \& Li $\gamma=1.16$). Thus, the
fireball becomes sub-relativistic at $t>100$~d and the flux scaling
with time, $f\propto t^{-1.4}$, which applies for $\gamma\gg1$,
changes to the sub-relativistic scaling $f\propto t^{-1.73}$, making
the discrepancy with the observed scaling even larger.

Second, let us consider the self-absorption frequency. Our data
clearly shows that the flux at 1.43~GHz is self absorbed up to at
least $t=100$~d. The time dependence of the self absorption frequency
in a wind model, $\nu_{ab}\propto\ t^{-3/5}$, therefore implies that
for this model to fit the late time radio data $\nu_{ab}\ge 7
(t/7d)^{-3/5}$~GHz is required (indeed the self-absorption frequency
for the parameters chosen in Chevalier \& Li is
$\nu_{ab}=6(t/1\,d)^{-3/5}$ GHz). This is in marked contradiction with
the value inferred from the data, $\nu_{ab}=3$~GHz (note that this
discrepancy in $\nu_{ab}$ implies a factor of $\simeq4$ discrepancy
between model and observations in the 1.43~GHz to 8.46~GHz flux
ratio).

\section{Discussion and Conclusions}

In this paper we present the 450-day radio afterglow curve of \grb.
This is by far the longest light curve obtained for any GRB.  We also
present a comprehensive model to explain the entire light curve.  In
our model, the fireball is a wide-angle jet which gradually expands
sideways and becomes a spherical sub-relativistic shell by about day
100.  The spherical symmetry of the fireball and the sub-relativistic
expansion allow us to carry out calorimetry of the explosion,
unhindered by unknown geometry. To our knowledge, this represents the
first time that one has been able to carry out absolute calorimetry of
a GRB explosion.  Below we summarize the salient features of the model
and the principal results presented in this paper.

The comprehensive model we present has three regimes: (1) early
afterglow, $2 < t < 25$ d, (2) jet afterglow, $25 < t < 100$ d and (3)
spherical sub-relativistic afterglow.  Several authors, in addition to
us, have modeled the X-ray, optical and radio-afterglow observations
of the early afterglow (e.g. Waxman 1997, Wijers \&\ Galama 1999,
Granot et al. 1999, WKF98).  The modeling yield electron and magnetic
field equipartition factors of $\epsilon_e\sim \epsilon_B\sim 0.1$.
Estimates of the total explosion energy depend on the assumed fireball
geometry (spherical versus jet).
The high Lorentz factor hides the true geometry of the fireball. The
inferred isotropic afterglow energy is $\sim 10^{52}$ erg which is
comparable to the isotropic $\gamma$-ray energy release of
$3.4\times 10^{51}$ erg.

The spherical model which explains the early afterglow quite
nicely over-predicts the observed radio flux by day 25. We had earlier
proposed that this discrepancy could be explained by invoking a jet
geometry for the fireball (WKF98). To our knowledge this was the first
explicit suggestion for a jet in a GRB. 

We argue that the jet starts significant sideways expansion 
at $t\sim25$~d, and that by
day 100, the fireball becomes essentially a spherically symmetric
sub-relativistically expanding shell. The primary observational
evidence supporting the transition to the sub-relativistic regime is 
relation between power law temporal and spectral indices, which is 
inconsistent with relativistic expansion, and consistent with 
sub-relativistic expansion.

The radio observations presented here of the sub-relativistic phase of 
fireball expansion
are unique. In this phase, the geometry of the fireball is no longer
an issue, and the fireball dynamics is described by the
well known Sedov-von Neumann-Taylor self-similar solution. Thanks to
the extensive radio data (multi-band and long-term monitoring) we are
able to deduce three out of the four fundamental parameters of the
synchrotron emitting fireball shell: 
$B_0$, the magnetic field strengths; $n_i$, the
ambient density; $r_0$, the radius of the shell; and $\gamma_0$, the
Lorentz factor of the emitting electrons.  Leaving the radius as the
free parameter we estimate the total energy of the emitting
particles (including the energy in magnetic fields).  We then use the
constraint that this energy must not exceed the thermal energy of the
a Sedov-von Neumann-Taylor expanding shell.  
This constraint then yields the size,
and hence the total energy, of the expanding shell.

We find the following: $E_0\sim 5\times 10^{50}$ erg and $n_i\sim 0.5$
cm$^{-3}$. Furthermore, we find that the electrons and magnetic field
are close to equipartition with $\epsilon_e\sim \epsilon_B \sim 0.5$.
Confidence in this ``absolute'' calorimetry is obtained when we note
that the inferred size of the shell is comparable to that measured
from interstellar scintillations (during the early phase).
Furthermore, our model for $t>100$~d predicts that the transition to
sub-relativistic regime takes place at $t\sim100$, as is indeed
observed (see above).

Without any doubt, the most important result of this paper is the
estimate of the total energy released in the explosion.  This energy,
$E_0\sim 5\times 10^{50}$ erg is well below $10^{52}$ erg, determined
from early afterglow observations and the isotropic $\gamma$-ray
energy release of $3.4\times 10^{51}$ erg. Thus the energy release in
\grb\ may not be much larger than that in ordinary SNe.

\acknowledgements

SRK and DAF are indebted to Barry Clark for his generous scheduling of
the VLA for this long project. DAF thanks J. Katz and B. Paczy\'nski
for useful discussions. SRK's research is supported by NSF and NASA.
EW's research is supported by AEC 38/99 and by BSF 9800343.

\appendix

\section{A Sub-Relativistic Model}
\label{sec:NRModel}

\subsection{Model assumptions}

Since the fireball expands at a sub-relativistic velocity in our
model, we assume to leading order that the radio flux observed at any
given time is emitted by a static shell of radiating electrons. We
denote the shell radius by $r$. During sub-relativistic expansion, the
dynamics of the shell are described by the self-similar Sedov--von
Neumann--Taylor solution. The shell radius and its velocity $\beta c$
scale with time as
\begin{equation}
r=r_0(t/t_0)^{2/5}, \quad \beta=\beta_0(t/t_0)^{-3/5}\,.
\label{eq:r_t}
\end{equation}
Here $t_0$ is some (arbitrary) reference time, and the $0$ subscripts
denote parameter values at $t=t_0$. In what follows we use the
normalization $t_0=100$~d.  The compressed, shock heated, material
behind the shock occupies a thin shell of width $r/\eta$, with
$\eta\approx 10$. We assume that plasma parameters are uniform within
the shell.

Electrons are assumed
to be shock accelerated to a power-law energy distribution,
$dn/d\gamma\propto\gamma^{-p}$ for $\gamma>\gamma_m$. We choose
$p=2.2$, as required by the high frequency observations. The magnetic field
energy density, $B^2/8\pi$, and electron energy density,
$(p-1)\gamma_m n m_e c^2/(p-2)$ are assumed to constitute fixed
(time-independent) fractions of the shock thermal energy density.
Under these assumptions, and recalling that the thermal energy density
is proportional to $\beta^2$,
we have
\begin{equation}
B=B_0 (t/t_0)^{-3/5}, \quad \gamma_m=\gamma_{m,0} (t/t_0)^{-6/5}.
\label{eq:time_dependence}
\end{equation}

\subsection{Detailed spectrum}

The synchrotron power per unit frequency $\nu$ emitted by a single electron
of Lorentz factor $\gamma$ is given by (Rybicki \& Lightman 1979),
\begin{equation}
P(\nu,\gamma)={e^3 B\over m_e c^2} F\left[{\nu \over
\nu_c(B,\gamma)}\right],
\label{eq:single_e}
\end{equation}
where $e$ and $m_e$ are the electron charge and mass, and
\begin{equation}
\nu_c\equiv \gamma^2\left({eB\over 2\pi m_ec}\right).
\label{eq:frequency}
\end{equation}
The function $F(x)$ describes the synchrotron power spectrum (Rybicki \&
Lightman 1979), averaged over an isotropic distribution of pitch angles.
To obtain the observed flux at a distance $d$ one needs to integrate over
the the angular
distribution of the intensity $I_\nu$ at the surface of the shell,
$f_\nu=2\pi(r/d)^2\int_0^1{\rm d}\cos\theta I_\nu(\theta)\cos\theta$.
For a thin, $\eta\gg1$, uniform shell, this integral is well
approximated by
\begin{equation}
f_\nu=\left({1-e^{-\tau_\nu}\over \tau_\nu}\right)
4\pi\left({r\over d}\right)^2{r\over\eta}\ j_\nu,
\label{eq:flux}
\end{equation}
provided the effective optical depth is chosen as
$\tau_\nu\equiv4\alpha_\nu(r/\eta)$.
Here $j_\nu$ is the synchrotron emissivity, $d=d_L(z)/(1+z)^{1/2}$,
and $\alpha_\nu$ is the
synchrotron self-absorption coefficient.

%The self-absorption coefficient is given by (Rybicki \& Lightman 1979),
%\begin{equation}
%\alpha_\nu=-{1\over 8\pi m_e\nu^2}\int d\gamma^\prime
%P(\nu,\gamma^\prime)\gamma^{\prime 2}
%{\partial \over \partial\gamma^\prime}
%\left( {1\over \gamma^{\prime 2}}{dn\over d\gamma^\prime}\right).
%\label{eq:optical_depth}
%\end{equation}

For a power-law electron distribution we obtain
\begin{equation}
f_\nu= a (t/t_0)^{11/10}
(1-e^{-\tau_\nu})[(1+z)\nu]^{5/2} f_3\left({\nu\over \nu_m}\right)
f_2^{-1}\left({\nu\over \nu_m}\right),
\label{eq:totalflux}
\end{equation}
\begin{equation}
\tau_\nu\equiv b (t/t_0)^{1-(3p/2)}
[(1+z)\nu]^{-(p+4)/2} f_2\left({\nu\over \nu_m}\right).
\label{eq:tau}
\end{equation}
Here
\begin{equation}
\nu_m\equiv{\nu_c(\gamma=\gamma_m)\over1+z}={\nu_0\over1+z}(t/t_0)^{-3},
\label{eq:nu_m}
\end{equation}
the functions
\begin{equation}
f_l(x)\equiv \int_0^x dy F(y) y^{(p-l)/2},
\label{eq:f_l}
\end{equation}
and the constants
\begin{equation}
\nu_0\equiv\frac{1}{2\pi}\gamma_0^2{eB_0\over m_ec},
\label{eq:nu0}
\end{equation}
\begin{equation}
a\equiv \frac{2\pi}{2+p} m_e\left({r_0\over d}\right)^2
\left({2\pi m_e c\over e B_0}\right)^{1/2},
\label{eq:a}
\end{equation}
\begin{equation}
b\equiv \frac{(2+p)(p-1)}{4\pi\eta} \gamma_0^{p-1} n
\left({e^3B_0r_0\over m_e^2 c^2}\right)
\left({eB_0\over 2\pi m_e c}\right)^{p/2}.
\label{eq:b}
\end{equation}

The model spectrum is determined by three parameters, $a$, $b$
and $\nu_0$. These parameters, in turn, are determined by four model
parameters, $B_0$, $r_0$, $\gamma_0$ and $n$.

\subsection{Simple scalings}

We define the self-absorption frequency $\nu_{ab}$ by $\tau_\nu(\nu=\nu_{ab})=1$.
For $\nu_{ab}\ll\nu_m$, i.e.
when the self-absorption frequency is much smaller than the characteristic
synchrotron radiation
frequency of the lowest energy electrons, $f_\nu$ peaks
at $\nu_p\approx\nu_m$. The peak frequency and flux scale as
\begin{equation}
\nu_p\approx\nu_m\propto t^{-3},\quad f_p\approx f_{\nu_m}\propto t^{3/5},
\label{eq:f_p1}
\end{equation}
and $\nu_{ab}$ scales as
\begin{equation}
\nu_{ab}\propto t^{6/5}.
\label{eq:nu_A1}
\end{equation}
The temporal and frequency dependence of $f_\nu$ is

\begin{equation}
f_\nu\propto
   \cases{
     \nu^2 t^{-2/5}                 & $\nu\ll\nu_{ab}$,\cr
     \nu^{1/3} t^{8/5}              & $\nu_{ab}\ll\nu\ll\nu_m$,\cr
     \nu^{-\beta} t^{-3\beta+(3/5)} & $\nu_{m}\ll\nu$,\cr
         }
\label{eq:f_nu1}
\end{equation}
where $\beta=(p-1)/2$.

For $\nu_{ab}\gg\nu_m$, $f_\nu$ peaks at frequency $\nu_p\approx\nu_{ab}$.
The optical depth at the peak frequency,
$\tau_p\equiv\tau_\nu(\nu=\nu_p)$, satisfies
\begin{equation}
\frac{5}{p+4}\left(1-e^{-\tau_p}\right)=\tau_p e^{-\tau_p}.
\label{eq:tau_p}
\end{equation}
For $p=2$ this implies $\tau_p=0.35$, $\nu_p=1.42\nu_{ab}$, and for $p=3$
we have $\tau_p=0.64$, $\nu_p=1.14\nu_{ab}$.
In this case,
\begin{equation}
\nu_p\approx\nu_{ab}\propto t^{-3(p-2/3)/(p+4)},\quad
f_p\propto t^{(3/5)-7(p-1)/(p+4)},
\label{eq:f_p2}
\end{equation}
and the temporal and frequency dependence of $f_\nu$ is
\begin{equation}
f_\nu\propto
   \cases{
     \nu^2 t^{13/5}        & $\nu\ll\nu_m$,\cr
     \nu^{5/2} t^{11/10}      & $\nu_m\ll\nu\ll\nu_{ab}$,\cr
     \nu^{-\beta} t^{-3\beta+(3/5)} & $\nu_{ab}\ll\nu$.\cr
         }
\label{eq:f_nu2}
\end{equation}

In our analysis we have so far neglected the effects of synchrotron
cooling of electrons. Synchrotron cooling is negligible for electrons
radiating at radio frequencies, since $\nu_c\approx5\times10^{13}$~Hz
at $t=100$~d (see \S5), but may be important for electrons radiating
at higher frequencies. The electron synchrotron cooling time is
proportional to $1/\gamma B^2$, where $\gamma$ is the electron Lorentz
factor, while the shell expansion time is proportional to $t$. Thus,
the Lorentz factor $\gamma_c$ of electrons for which the cooling time
is comparable to the expansion time scales as $\gamma_c\propto1/B^2t$,
and the frequency $\nu_c$, beyond which electron cooling affects the
observed radiation, scales as
\begin{equation}
\nu_c\propto\gamma_c^2B\propto B^{-3}t^{-2}\propto t^{-1/5}.
\label{eq:nu_c}
\end{equation}
At frequencies higher than $\nu_c$, electron cooling steepens the spectrum
and
\begin{equation}
f_\nu\propto\nu^{-\beta-(1/2)}t^{-3\beta+(1/2)}, \quad \nu>\nu_c.
\label{eq:f_nuc}
\end{equation}
Note that for $p=2$, $f_\nu\propto t^{-1}$ (at $\nu>\nu_c$) for both
relativistic and sub-relativistic expansion.

\clearpage
\begin{deluxetable}{lrcrc}
%\footnotesize
\scriptsize
\tablecaption{1.43 GHz Observations of GRB 970508\label{20obs}}
\tablewidth{0in} \tablehead{ \colhead{Date} &
\colhead{$\Delta{t}$} & \colhead{F$_{1.4}$\tablenotemark{a}} &
\colhead{$\sigma_{1.4}$} &
\colhead{Array} \\
\colhead{(UT)} & \colhead{(days)} & \colhead{($\mu$Jy)} &
\colhead{($\mu$Jy)} & \colhead{Config.\tablenotemark{b}}}\tablecolumns{5}
\startdata
1997 May  \phantom{1}9.05  &        0.15   & \nodata       & 45  & B \nl
1997 May  \phantom{1}9.84  &        0.93   & \nodata       & 89  & B \nl
1997 May  15.09 &        6.19   & 100   & 32  & B \nl
1997 May  18.83 &       9.93    & \nodata       & 63  & B \nl
1997 Jun. \phantom{1}1.00  &       23.10   & \nodata       & 52  & CnB \nl
1997 Jun. \phantom{1}2.32  &       24.41   & 135   & 45  & CnB \nl
1997 Jun. 12.83 &       34.93   & 130   & 20  & CnB \nl
1997 Jul. 18.62 &       70.72   & 305   & 48  & C \nl
1997 Jul. 28.94 &       81.03   & 485   & 45  & CS \nl
1997 Aug. 19.02 &       102.11  & 250   & 70  & CS \nl
1997 Aug. 22.52 &       105.62  & 230   & 59  & CS \nl
1997 Sep. \phantom{1}3.47  &       117.56  & 206   & 63  & CS \nl
1997 Sep. \phantom{1}8.42  &       122.51  & 216   & 67  & CS \nl
1997 Sep. 12.68 &       126.78  & 333   & 65  & CS \nl
1997 Sep. 26.35 &       140.45  & \nodata       & 83  & DnC \nl
1997 Oct. \phantom{1}4.63  &       148.73  & 217   & 105 & DnC \nl
1997 Oct. 13.22 &       157.32  & 381   & 127 & DnC \nl
1997 Oct. 21.39 &       165.48  & 250   & 66  & DnC \nl
1997 Nov. \phantom{1}8.09  &       183.19  & 338   & 63  & D \nl
1997 Nov. 11.33 &       186.42  & 497   & 51  & D \nl
1997 Nov. 13.97 &       189.06  & 376   & 93  & D \nl
1997 Nov. 17.65 &       192.75  & \nodata       & 93  & D \nl
1997 Nov. 19.46 &       194.55  & 303   & 53  & D \nl
1997 Nov. 22.46 &       197.56  & 199   & 62  & D \nl
1997 Nov. 23.05 &       198.14  & 271   & 74  & D \nl
1997 Nov. 24.96 &       200.05  & 208   & 77  & D \nl
1997 Nov. 26.17 &       201.26  & 155   & 43  & D \nl
1997 Nov. 27.12 &       202.21  & 170   & 50  & D \nl
1997 Nov. 28.40 &       203.50  & \nodata       & 62  & D \nl
1997 Nov. 30.32 &       205.42  & 184   & 42  & D \nl
1997 Dec. \phantom{1}1.40  &       206.49  & \nodata       & 71  & D \nl
1997 Dec. \phantom{1}6.65  &       211.75  & \nodata       & 68  & D \nl
1997 Dec. 11.16 &       216.25  & \nodata       & 64  & D \nl
1998 Feb. 23.44 &       290.53  & 274   & 27  & A \nl
1998 Mar. 18.06 &       313.16  & 434   & 24  & A \nl
1998 Apr. 27.66 &       353.76  & 249   & 60  & A \nl
1998 May  \phantom{1}2.52  &       358.62  & 204   & 30  & A \nl
1998 Jul. 19.78 &       436.89  & \nodata       & 32  & B \nl
1998 Jul. 28.73 &       445.83  & \nodata       & 33  & B \nl
\enddata
\tablenotetext{a}{When no flux density measurement is given
it can be assumed that the source was not detected above the
3$\sigma_{1.4}$ level.}
\tablenotetext{b}{The 27 antennas of the VLA can be placed into one of
  four ``standard'' configurations (A, B, C and D) giving a
  synthesized beamsize (i.e.~resolution) of 1.2\arcsec\ in A-array and
  increasing in multiples of 3.3 to the D-array resolution of
  44\arcsec. Hybrid arrays such as DnC and CnB have increased
  resolution in the north-south direction, while the CS array is a
  special C-array hybrid designed for increased sensitivity to
  extended emission.}
\end{deluxetable}

\clearpage

\begin{deluxetable}{lrcrcrc}
%\footnotesize
\scriptsize
\tablecaption{4.86 and 8.46 GHz Observations of GRB 970508\label{tab:63obs}}
\tablewidth{0in} \tablehead{ \colhead{Date} &
\colhead{$\Delta{t}$} &
\colhead{F$_{4.86}$} &
\colhead{$\sigma_{4.86}$} &
\colhead{F$_{8.46}$} &
\colhead{$\sigma_{8.46}$} &
\colhead{Array} \\
\colhead{(UT)} &
\colhead{(days)} &
\colhead{($\mu$Jy)} &
\colhead{($\mu$Jy)} &
\colhead{($\mu$Jy)} &
\colhead{($\mu$Jy)} &
\colhead{Config.\tablenotemark{a}}}\tablecolumns{5}
\startdata
1997 May 13.96  & 5.06  & \nodata & \nodata & 430 & 25 & B \nl
1997 May 15.13  & 6.23  & 330 & 33 & 610 &  33 & B \nl
1997 May 16.49  & 7.59  & \nodata & \nodata  & 520 &  18 & B \nl
1997 May 16.79  & 7.89  & \nodata & \nodata & 602 & 24 & V \nl
1997 May 17.30  & 8.40  & \nodata & \nodata & 604 & 56 & V \nl
1997 May 18.85  & 9.95  & 480 & 64 & 610 &  51 & B \nl
1997 May 21.00  & 12.10 & \nodata & \nodata  & 1270 &  120 & V \nl
1997 May 21.13  & 12.23 & \nodata & \nodata & 930 &  120 & V \nl
1997 May 22.10  & 13.20 & 410 & 47 & 570 &  42 & B \nl
1997 May 22.49  & 13.59 & 360 & 31 & 880 &  33 & B \nl
1997 May 23.00  & 14.10 & \nodata & \nodata & 1260 &  100 & V \nl
1997 May 23.13  & 14.23 & \nodata & \nodata & 960 &  100 & V \nl
1997 May 24.48  & 15.58 & \nodata & \nodata & 550 &  44 & B \nl
1997 May 24.96  & 16.06 & \nodata & \nodata & 470 &  43 & B \nl
1997 May 25.85  & 16.95 & \nodata & \nodata & 480 &  45 & B \nl
1997 May 27.67  & 18.77 & 770 & 54 & 500 &  49 & B \nl
1997 May 28.10  & 19.20 & \nodata & \nodata & 538 &  100 & V \nl
1997 May 28.23  & 19.33 & \nodata & \nodata & 835 &  100 & V \nl
1997 May 29.10  & 20.20 & 1350 & 55 & 1200 &  47 &  CnB \nl
1997 May 30.95  & 22.05 & 740 & 46 & 810 &  48 & CnB \nl
1997 May 31.99  & 23.09 & 550 & 43 & 290 &  46 & CnB \nl
1997 Jun. 2.29  & 24.39 & 650 & 44 & 720 &  62 & CnB \nl
1997 Jun. 2.41  & 24.51 & \nodata & \nodata & 940 &  27 & CnB \nl
1997 Jun. 2.50  & 24.60 & \nodata & \nodata & 960 &  40 & V \nl
1997 Jun. 2.90  & 25.00 & 540 & 46 & 760 &  45 & CnB \nl
1997 Jun. 3.94  & 26.04 & 530 & 51 & 670 &  43 & CnB \nl
1997 Jun. 5.97  & 28.07 & 850 & 36 & 690 &  28 & CnB \nl
1997 Jun. 9.74  & 31.84 & 470 & 38 & 530 &  38 & CnB \nl
1997 Jun. 12.99 & 35.09 & 640 & 43 & 600 &  39 & CnB \nl
1997 Jun. 14.73 & 36.83 & 590 & 69 & 460 &  66 & CnB \nl
1997 Jun. 17.92 & 40.02 & 400 & 49 & 560 &  40 & CnB \nl
1997 Jun. 18.96 & 41.06 & 640 & 55 & 630 &  63 & CnB \nl
1997 Jun. 20.86 & 42.96 & 550 & 36 & 800 &  32 & CnB \nl
1997 Jun. 22.93 & 45.03 & 425 & 43 & 645 &  38 & CnB \nl
1997 Jun. 26.63 & 48.73 & 610 & 70 & 680 &  46 & CnB \nl
1997 Jun. 28.68 & 50.78 & 500 & 34 & 565 &  36 & C \nl
1997 Jul. 2.36  & 54.46 & 805 & 38 & 805 &  43 & C \nl
1997 Jul. 4.59  & 56.69 & 605 & 50 & 710 &  49 & C \nl
1997 Jul. 11.10 & 63.20 & 630 & 32 & 725 &  40 & C \nl
1997 Jul. 14.61 & 66.71 & 585 & 48 & 680 &  46 & C \nl
1997 Jul. 15.33 & 67.43 & 550 & 31 & 630 &  37 & C \nl
1997 Jul. 18.44 & 70.54 & 505 & 62 & 575 &  61 & C \nl
1997 Jul. 18.64 & 70.74 & 535 & 43 & 515 &  41 & C \nl
1997 Jul. 22.55 & 74.65 & 385 & 51 & 440 &  51 & CS \nl
1997 Jul. 30.46 & 82.56 & 360 & 70 & 455 &  85 & CS \nl
1997 Aug. 4.01  & 87.11 & 370 & 38 & 380 &  42 & CS \nl
1997 Aug. 5.35  & 88.45 & 550 & 61 & 445 &  50 & CS \nl
1997 Aug. 7.23  & 90.33 & 515 & 37 & 430 &  39 & CS \nl
1997 Aug. 11.26 & 94.36 & 615 & 37 & 640 &  40 & CS \nl
1997 Aug. 15.13 & 98.23 & 770 & 57 & 565 &  49 & CS \nl
1997 Aug. 16.62 & 99.72 & 610 & 55 & 400 &  60 & CS \nl
1997 Aug. 18.95 & 102.05 & 500 & 42 & 300 &  35 & CS \nl
1997 Aug. 22.50 & 105.60 & 710 & 52 & 470 &  44 & CS \nl
1997 Aug. 31.82 & 114.92 & \nodata & \nodata & 315 & 70 & DnC \nl
1997 Sep. 3.45  & 117.55 & 425 & 57 & 355 &  47 & DnC \nl
1997 Sep. 4.71  & 118.81 & \nodata & \nodata & 370 &  70 & DnC \nl
1997 Sep. 8.40  & 122.50 & \nodata & \nodata & 350 &  47 & DnC \nl
1997 Sep. 12.64 & 126.74 & 400 & 41 & 370 &  39 & DnC \nl
1997 Sep. 26.36 & 140.46 & 380 & 40 & 330 &  45 & DnC \nl
1997 Oct. 4.62  & 148.72 & 420 & 44 & 310 &  56 & DnC \nl
1997 Oct. 13.23 & 157.33 & 350 & 40 & 280 &  37 & DnC \nl
1997 Oct. 21.40 & 165.50 & 300 & 34 & 240 &  34 & DnC \nl
1997 Oct. 26.56 & 170.66 & 280 & 31 & 300 &  37 & DnC \nl
1997 Nov. 6.19  & 181.29 & 385 & 32 & 240 &  24 & D \nl
1997 Nov. 19.15 & 194.25 & 325 & 37 & 240 &  30 & D \nl
1997 Dec. 9.14  & 214.24 & 180 & 35 & 170 &  30 & D \nl
1998 Jan. 2.08  & 238.18 & 220 & 35 & 160 &  30 & D \nl
1998 Jan. 18.90 & 255.00 & 275 & 53 & \nodata & \nodata  & D \nl
1998 Jan. 20.98 & 257.08 & 235 & 41 & 150 &  36 & D \nl
1998 Jan. 27.77 & 263.87 & 220 & 50 & \nodata &  50 & D \nl
1998 Jan. 31.29 & 267.39 & \nodata & 80 & \nodata &  50 & D \nl
1998 Feb. 8.04  & 275.14 & 200 & 24 & 95 &  24 & A \nl
1998 Mar. 14.09 & 309.19 & 165 & 17 & 105 &  15 & A \nl
1998 May 13.52  & 369.62 & 125 & 28 & \nodata &  25 & A \nl
1998 Jun. 21.83 & 408.93 & 104 & 44 & \nodata &  37 & BnA \nl
\enddata
\tablenotetext{a}{Letter designations for the array configurations are
  explained in the table notes of Table \ref{20obs}. A ``V''
  denotes flux density measurements obtained using the VLBA by Taylor
  et al. (1997).}
\end{deluxetable}

\clearpage
\begin{deluxetable}{ccccl}
%\footnotesize
\scriptsize
\tablecaption{The Transition to the Power-Law Regime\label{tab:nuobs}}
\tablewidth{0in} \tablehead{\colhead{$\nu_m$} &
\colhead{$t_m$} & \colhead{F$_{\nu_m}$} & \colhead{$\sigma_{rms}$} &
\colhead{References} \\
\colhead{(GHz)} & \colhead{(days)} & \colhead{($\mu$Jy)} &
\colhead{($\mu$Jy)} &
\colhead{$\phantom{ref}$}}\tablecolumns{5}
\startdata
86.2   & 10-14     & 1620 & 250 & Bremer et al. (1998)   \nl
15.0   & 8-13      &  660 & 110 & Pooley \& Green (1997) \nl
8.46   & $\sim$90  &  630 & 107 & this paper \nl
4.86   & $\sim$90  &  555 & 108 & this paper \nl
1.43   & 310-490   &  300 & 150 & this paper \nl
\enddata
\end{deluxetable}
% Table notes. wlsq fiit gives 0.36+/-0.06. lsq gives 0.40+/-0.04

\clearpage
\begin{deluxetable}{cccl}
%\footnotesize
\scriptsize
\tablecaption{Spectral and Temporal Power-Law Indices\label{tab:abobs}}
\tablewidth{0in} \tablehead{\colhead{Time} &
\colhead{$\alpha$\tablenotemark{a}}
& \colhead{$\beta$\tablenotemark{a}} & \colhead{References} \\
\colhead{Range (days)} & \colhead{} & \colhead{} &
\colhead{$\phantom{ref}$}}\tablecolumns{5}
\startdata
2-4     & $-0.9\pm{0.1}$     & $-0.65\pm{0.3}$  & Djorgovski et al. (1997) \nl
2-25    & $-1.13\pm{0.04}$   & $-1$             & Pian et al. (1998) \nl
2-120   & $-1.191\pm{0.024}$ & $-1.11\pm{0.06}$ & Galama et al. (1998c) \nl
2-90    & $-1.171\pm{0.012}$ & $-1.1$           & Sokolov et al. (1998) \nl
2-100   & $-1.23\pm{0.04}$   & $-1.10\pm{0.08}$ & Zharikov et al. (1998) \nl
90-300  & $-1.14\pm{0.06}$    & $-0.50\pm{0.06}$   & this paper \nl
%90-300  & $-1.2\pm{0.06}$    & $-0.6\pm{0.2}$   & this paper \nl
\enddata
\tablenotetext{a}{The flux density $f_\nu\propto{t}^\alpha\nu^\beta$.}
\end{deluxetable}

\clearpage

%\begin{figure}
%\centerline{\psfig{file=w2.ps,height=6cm}}
%\centerline{\psfig{file=w3.ps,height=6cm}}
%\centerline{\psfig{file=w4.ps,height=6cm}}
%\caption[]{Model light curves compared to data.
%\label{fig:ewfits}}
%\end{figure}

\end{document}